\documentclass[pre,twocolumn,amsmath,eqsecnum,fleqn,showpacs,preprintnumbers,twoside]{revtex4-1}

\usepackage{amsfonts}
\usepackage{amssymb}
\usepackage{graphicx}
\usepackage{stmaryrd}
\usepackage{color}

\def\Rc{R}

\def\cpt{c}

\def\RSGP{S}

\begin{document}

%%%%%%%%%%%%%%%%%%%%%%%%%%%%%%%%%%%%%%%%%%%%%%%%%%%%%%%%%%%%%%%%%%%%%
\title{
Ballistic, diffusive, and localized transport in surface-disordered
systems:\\Two-mode waveguide
}
%%%%%%%%%%%%%%%%%%%%%%%%%%%%%%%%%%%%%%%%%%%%%%%%%%%%%%%%%%%%%%%%%%%%%

%%%%%%%%%%%%%%%%%%%%%%%%%%%%%%%%%%%%%%%%%%%%%%%%%%%%%%%%%%%%%%%%%%%%%
\author{M.~Rend\'{o}n}
\email{mrendon@ece.buap.mx}
\affiliation{Facultad de Ciencias de la Electr\'{o}nica,
         Universidad Aut\'{o}noma de Puebla,
         Puebla, Pue., 72570, M\'{e}xico}

\author{N.~M.~Makarov}
\email{makarov@siu.buap.mx}
\affiliation{Instituto de Ciencias, Universidad Aut\'{o}noma
         de Puebla, \\ Priv. 17 Norte No. 3417, Col. San Miguel
         Hueyotlipan, Puebla, Pue., 72050, M\'{e}xico}

\author{F.~M.~Izrailev}
\email{izrailev@sirio.ifuap.buap.mx}
\affiliation{Instituto de F\'{\i}sica, Universidad Aut\'{o}noma
         de Puebla, \\
         Apartado Postal J-48, Puebla, Pue., 72570, M\'{e}xico}
%%%%%%%%%%%%%%%%%%%%%%%%%%%%%%%%%%%%%%%%%%%%%%%%%%%%%%%%%%%%%%%%%%%%%

\date{\today}

%%%%%%%%%1%%%%%%%%%2%%%%%%%%%3%%%%%%%%%4%%%%%%%%%5%%%%%%%%%6%%%%%%%%%7%
\begin{abstract}
This paper presents an analytical study of the coexistence of different 
transport regimes in quasi-one-dimensional surface-disordered waveguides 
(or electron conductors). To elucidate main features of surface scattering, 
the case of two open modes (channels) is considered in great detail. Main 
attention is paid to the transmission in dependence on various parameters 
of the model with two types of rough-surface profiles (symmetric and 
antisymmetric). It is shown that depending on the symmetry, basic 
mechanisms of scattering can be either enhanced or suppressed. As a 
consequence, different transport regimes can be realized. Specifically, in the 
waveguide with symmetric rough boundaries, there are ballistic, localized 
and coexistence transport regimes. In the waveguide with antisymmetric 
roughness of lateral walls, another regime of the diffusive transport can arise. 
Our study allows to reveal the role of the so-called square-gradient scattering 
which is typically neglected in literature, however, can give a strong impact to 
the transmission.
\end{abstract}
%%%%%%%%%%%%%%%%%%%%%%%%%%%%%%%%%%%%%%%%%%%%%%%%%%%%%%%%%%%%%%%%%%%%%


\maketitle

%%%%%%%%%%%%%%%%%%%%%%%%%%%%%%%%%%%%%%%%%%%%%%%%%%%%%%%%%%%%%%%%%%%%%%%%%%%%%%%
\section{Introduction}
\label{sec:Intro}
%%%%%%%%%%%%%%%%%%%%%%%%%%%%%%%%%%%%%%%%%%%%%%%%%%%%%%%%%%%%%%%%%%%%%%%%%%%%%%%

The transport of quantum quasi-particles through thin metal films and 
semiconductor nano-structures, such as nanowires and strips, supperlattices 
and quantum-well systems, has been a hot topic since more than four 
decades~\cite{Chopra_1969,AbrRyz_AdvPhys_1978, 
Lnd_PhysScr_1992,Dtt_1995,DttHngIngKrmSchZwr_1998,AkkMnt_2007}. 
Nowadays, it is known that transport properties of conductors with extremely 
small cross sections, are substantially controlled by the scattering of electrons at 
random inhomogeneities of the conductor 
boundaries~\cite{TrvAsh_1988,TkgFrr_JPhysCondMatt_1992,
KzbKrk_JPhysCondMatt_1993,MyrStp_PhysRevB_1995,
MkrMrzYmp_PhysRevB_1995}.
Many of transport properties of mesoscopic guiding systems can be modeled as 
surface-disordered waveguides.

Recent numerical studies of quasi-one-dimensional (quasi-1D) 
surface-disordered systems~\cite{GrcTrrSnzNto_1998,SncFrlYrkMrd_1998,
SncFrlMrdYrk_1999} have revealed a key difference from the standard models 
with bulk random potentials. Specifically, it was found that transport properties of 
waveguides with surface scattering essentially depend on multiple characteristic 
lengths, in contrast with the conductors with bulk scattering for which the 
single-parameter scaling~\cite{AbrAndLccRmk_PhysRevLett_1979} occurs. 
According to the single-parameter scaling, all scattering properties of finite 
conductors with bulk random potentials are determined by a single parameter: 
the ratio of the conductor size to the localization length found for the 
corresponding systems of infinite size. However, the scattering properties of 
surface-disordered waveguides are determined by a non-isotropic character of 
scattering in the ``channel space.''

For surface-corrugated waveguides with a large number of conducting channels 
$N_d\gg1$, it was 
emphasized~\cite{IzrMkr_JPhysA_2005,IzrMkrRnd_PRB_2005,
IzrMkrRnd-PRB_2006,RndIzrMkr-MSMW-2007,RndIzrMkr-PRB-2007,
RndIzrMkr-MMET-2008} that the mode-attenuation length 
$L_n$ of the $n$th propagating mode (or mean free path of the $n$th conducting 
channel) displays a rather strong dependence on the mode index $n$. 
Specifically, the larger the number $n$ the smaller the corresponding 
mode-attenuation length and, as a consequence, the stronger is the scattering of 
this mode into the others. As a result, there emerges the \emph{hierarchy} of 
mode-attenuation lengths~\cite{IzrMkr_JPhysA_2005}:
\begin{equation}\label{eq:hierarchy}
L_{N_d}<L_{N_d-1}<\ldots<L_2<L_1.
\end{equation}
The smallest length $L_{N_d}$ belongs to the highest (last) channel with the 
mode index $n=N_d\gg1$, while the largest length, $L_1$, corresponds to the 
lowest (first) channel with $n=1$. Because of this hierarchy of lengths, a very 
important phenomenon can arise, that is the \emph{coexistence} of ballistic, 
diffusive, and localized transport~\cite{SncFrlYrkMrd_1998, SncFrlMrdYrk_1999}.
For example, for a conductor of length $L$ which is inside the set of lengths 
$L_n$,
\begin{equation}\label{eq:LHierarchy}
L_{N_d}<\ldots<L<\ldots<L_1,
\end{equation}
the lowest modes can be in ballistic regime, however, the intermediate and 
highest modes can exhibit the diffusive and localized behavior, respectively.

Accordingly, for quasi-1D surface-disordered waveguides one can state the 
following:
(i) The average mode-transmittances $\langle T_M\rangle$, can be very different 
for different conducting channels.
(ii) All propagating modes are mixed because of inter-mode transitions, therefore, 
the transmittance $\langle T_M\rangle$ of any given $n$th mode depends on the 
scattering into all modes.
(iii) The total average transmittance, $\langle T\rangle$, of the waveguide 
contains the imprint of all the average mode-transmittances and, therefore, can 
exhibit an \emph{coexistence} transport in which the phenomenon of 
\emph{coexistence} of different mode-transport regimes takes place.

\begin{figure*}[ht!]
\includegraphics[width=12cm]{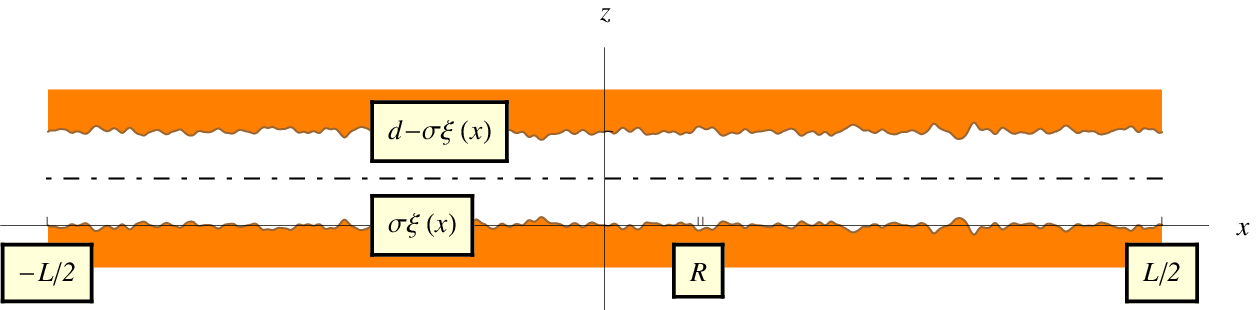}
\includegraphics[width=12cm]{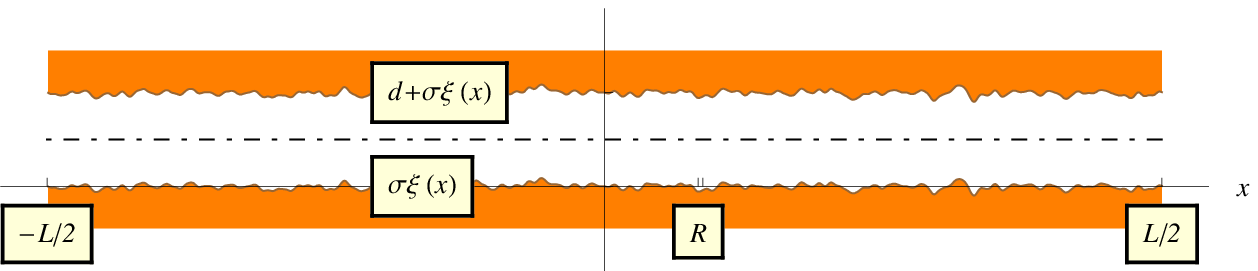}
\caption{\label{fig:SurfProfileWaveguides}
(Color online) One realization of symmetric (top) and antisymmetric (bottom)
surface-disordered waveguides, see in the text.
%Lengths in this picture are in scale with
% $\sigma/d=3/100$, $?R/d=5/100$ and $L/d=1190/100$.
}
\end{figure*}

In this paper the above statements are quantitatively validated by means of two 
special models of \emph{two-mode} waveguides with lateral symmetric rough 
boundaries (SRB) and antisymmetric rough boundaries (ASRB), see 
Fig.~\ref{fig:SurfProfileWaveguides}. The SRB waveguide represents a system in 
which the attenuation of the propagating modes is associated with the scattering 
within the same mode (intra-mode scattering). Comparatively, the attenuation in 
the ASRB waveguide arises due to the intra-mode scattering {\it and} due to the 
scattering of waves from one mode into the other (inter-mode scattering).

Although main attention is paid to the conduction of electrons, the results are also 
applicable to the propagation of classical waves. The latter problem has even a 
longer history, it naturally arises in the analysis of spectral and transport 
properties of optic fibers, acoustic and radio waveguides, remote sensing, 
shallow water waves, multilayered systems, photonic lattices and so 
forth~\cite{BssFks_1979,SntBrw_1986,Ogl_1991,FrlGrd_1992,OhlNvrOhl_1995}.

This paper is structured as follows. In Section~\ref{sec:Problem} the geometry 
and general statistical properties of the surface-disordered waveguides are 
described. Here the Green's function of the waveguides and standard linear 
response formula are outlined, which are used to obtain the transmittance (or 
dimensionless conductance). Also, the formulas for mode-attenuation lengths of 
waveguides are discussed in detail. In Sections \ref{sec:SRB} and 
\ref{sec:ASRB} the transport properties of the SRB and ASRB waveguides with 
two propagating modes are analyzed. Here the expressions for the mode and 
total average transmittances are derived. The transport regimes of the total 
average transmittance are then analyzed; one being the coexistence regime, in 
which the coexistence of different mode-transports emerges. Finally, the 
conclusions are presented in Section~\ref{sec:Conclusions}.

%%%%%%%%%%%%%%%%%%%%%%%%%%%%%%%%%%%%%%%%%%%%%%%%%%%%%%%%%%%%%%%%%%%%%%%%%%%%%%%
\section{General statement of the problem}
\label{sec:Problem}
%%%%%%%%%%%%%%%%%%%%%%%%%%%%%%%%%%%%%%%%%%%%%%%%%%%%%%%%%%%%%%%%%%%%%%%%%%%%%%%

%%%%%%%%%%%%%%%%%%%%%%%%%%%%%%%%%%%%%%%%%%%%%%%%%%%%%%%%%%%%%%%%%%%%%%%%%%%%%%%
\subsection{Surface-disordered waveguides}
%%%%%%%%%%%%%%%%%%%%%%%%%%%%%%%%%%%%%%%%%%%%%%%%%%%%%%%%%%%%%%%%%%%%%%%%%%%%%%%

This study considers open surface-disordered waveguides (or conducting wires) 
of length $L$ and average width $d$, stretched along the $x$- and $z$-axes, 
respectively; for quasi-1D geometry, naturally, $L\gg d$. Two different 
configurations of the opposite boundaries result in the symmetric and 
antisymmetric rough boundaries waveguides, see 
Fig.~\ref{fig:SurfProfileWaveguides}.
The SRB waveguide has inhomogeneities that are symmetric with respect to its 
straight central guiding axis (dot-dashed line in 
Fig.~\ref{fig:SurfProfileWaveguides}). 
Thus, the inhomogeneities give rise to the varying width of the SRB waveguide.
In contrast, the width of the ASRB waveguide stays constant along the 
waveguide despite the rough boundaries. Here the inhomogeneities are 
physically equivalent to a weak bending of waveguide.

The lower and upper surfaces of the waveguides are described, respectively, by 
the rough boundaries $z=\sigma\xi(x)$, and $z=d\mp \sigma\xi(x)$. The upper 
surface of the SRB (ASRB) waveguide is indicated by the minus (plus) sign of 
the symbol ``$\mp$''. Here $\sigma$ is the root-mean-square \emph{roughness 
height}, which is assumed to be identical for both boundaries. Hence, the 
waveguides occupy the region,
\begin{subequations}\label{eq:OccupiedRegion}
\begin{eqnarray}
-L/2 \leq & x & \leq L/2,\\
\sigma \xi(x)\leq & z & \leq d\mp \sigma \xi(x),
\end{eqnarray}
\end{subequations}
of the $(x,z)$-plane. The random function $\xi(x)$ describes the roughness of the 
boundaries and is assumed to be statistically homogeneous and isotropic, with 
the statistical properties of zero mean and unit variance,
\begin{subequations}\label{eq:xi-def}
\begin{eqnarray}
\label{eq:<xi>}
&&\langle\xi(x)\rangle=0,\\
\label{eq:<xi2>}
&&\langle\xi^2(x)\rangle=1,\\
\label{eq:<xixi>}
&&\langle\xi(x)\, \xi(x')\rangle=\mp {\cal W}(|x-x'|).
\end{eqnarray}
\end{subequations}
Here the angular brackets represent the statistical averaging over different 
realizations of the surface profile $\xi(x)$. The binary correlator ${\cal W}(x)$ 
decreases on the scale $\Rc$ with the normalization ${\cal W}(0)=1$, where 
$\Rc$ is the roughness correlation length. It should be reiterated that the minus 
(plus) sign of the correlator corresponds to the SRB (ASRB) waveguide. Since 
lower and upper boundaries have the same rough profile, the binary correlator 
\eqref{eq:<xixi>} describes the correlations within each boundary, as well as the 
cross-correlations between them.

%%%%%%%%%%%%%%%%%%%%%%%%%%%%%%%%%%%%%%%%%%%%%%%%%%%%%%%%%%%%%%%%%%%%%%%%%%%%%%%
\subsection{Transmittance and Green's function}
%%%%%%%%%%%%%%%%%%%%%%%%%%%%%%%%%%%%%%%%%%%%%%%%%%%%%%%%%%%%%%%%%%%%%%%%%%%%%%%

The transport properties of the disordered waveguide will be characterized by its 
\emph{transmittance} or, equally, by its dimensionless (in units of $e^2/\pi\hbar$) 
\emph{conductance} $T(L)$. Within the standard linear response theory, as 
indicated by R.~Kubo~\cite{Kubo_1957}, this quantity is expressed as
\begin{equation}\label{eq:T(L)-def}
\begin{split}
T(L)=&-\frac{4}{L^2}\int_{-L/2}^{L/2}dx\,dx'\,\int_{0}^{d}dz\,dz'\\
&\times\frac{\partial{\cal G}(x,x';z,z')}{\partial x}\,
\frac{\partial{\cal G}^*(x,x';z,z')}{\partial x'},
\end{split}
\end{equation}
where the asterisk ``$*$'' denotes complex conjugation. The retarded Green's 
function ${\cal G}(x,x';z,z')$ obeys the following boundary-value 
problem~\cite{RndIzrMkr-PRB-2007}:
\begin{subequations}\label{eq:GU_FlatBoundary}
\begin{eqnarray}
\label{eq:GU}
&&\left(\frac{\partial^2}{\partial x^2}+
\frac{\partial^2}{\partial z^2}+k^2\right){\cal G}(x,x';z,z')
\nonumber\\
&&-\hat{U}(x,z)\, {\cal G}(x,x';z,z')=\delta(x-x')\delta(z-z'), \nonumber \\ \\
\label{eq:FlatBoundary_0}
&&{\cal G}(x,x';z=0,z')=0,\\
\label{eq:FlatBoundary_d}
&&{\cal G}(x,x';z=d,z')=0.
\end{eqnarray}
\end{subequations}
Here $\delta(x)$ and $\delta(z)$ are the Dirac delta-functions. The wave number 
$k$ is equal to the Fermi wave number for the electrons within the isotropic 
Fermi-liquid model. For electromagnetic waves with both frequency $\omega$ 
and TE polarization, propagating through a waveguide with perfectly conducting 
walls, $k=\omega/c$.

The equations~\eqref{eq:GU_FlatBoundary} specify a Dirichlet boundary-value 
problem, however, with two remarkable features: the \emph{``bulk" scattering 
potential} $\hat{U}(x,z)$, and the \emph{flat} boundaries of the waveguide at 
$z=0$ and $z=d$. These features emerge after the coordinate transformation of 
the initial Dirichlet boundary-value problem, in which there is no ``bulk" potential 
while the scattering is only caused by the roughness of the boundaries. The idea 
of this method was first discussed by Migdal~\cite{Migdal_1977}, and has since 
been frequently used in theories of wave or electron surface-scattering. Details of 
the application of this method for ``flattening'' the rough boundaries, can be found 
in Ref.~\onlinecite{RndIzrMkr-PRB-2007}. The form of the potentials 
$\hat{U}(x,z)$ for the specific waveguides used in this study can be extracted 
from the aforementioned reference. It should be stressed that this potential can 
not be treated as a completely random potential and modeled, for example, by 
random matrices. A close inspection of this potential shows that random function 
$\xi(x)$ enters the potential in a rather complicated way, however, by no means, 
strong correlations between matrix elements can not be neglected.

For a further discussion we have to indicate that an approximate expression of 
the potential for the SRB waveguide contains several terms that can be joined 
together in three groups. These groups depend on $\sigma\xi(x)/d$, 
$\sigma\xi'(x)/d$, and $\sigma^2\xi'^2(x)/d^2$,
\begin{equation}\label{eq:SRB_U(x,z)}
\hat{U}(x,z)\approx \hat{U}\left[\frac{\sigma\xi(x)}{d}, 
\frac{\sigma\xi'(x)}{d},\frac{\sigma^2\xi'^2(x)}{d^2},z\right].
\end{equation}
An exact expression for the ASRB waveguide contains other terms, which, 
however, can be arranged in the groups that depend on $\sigma\xi'(x)/d$ and 
$\sigma^2\xi'^2(x)/d^2$,
\begin{equation}\label{eq:ASRB_U(x,z)}
\hat{U}(x,z)=\hat{U}\left[\frac{\sigma\xi'(x)}{d},\frac{\sigma^2\xi'^2(x)}{d^2},
z\right].
\end{equation}
Here the prime to the function $\xi(x)$ denotes a derivative with respect to $x$.

As is typical, after the application of the Green's theorem, the problem can be 
reformulated in the form of the Dyson equation. In order to perform the averaging 
of the Green's function, one can apply one of the well-known perturbative 
methods. For instance, the diagrammatic approach developed for 
surface-disordered systems~\cite{BssFks_1979}, or the technique developed in 
Ref.~\onlinecite{McGM84} may be equally utilized. Both methods take 
adequately into account the multiple scattering from rough boundaries and allow 
one to formulate a consistent perturbative approach with respect to the scattering 
potential. After straightforward but cumbersome calculations (see 
Ref.~\onlinecite{IzrMkrRnd-PRB_2006} for details), the following average 
Green's function is obtained,
\begin{equation}\label{eq:avGF}
\begin{split}
\langle {\cal G}(|x-x'|;z,z')\rangle=&\sum_{n=1}^{N_d}
\sin\left(\frac{\pi nz}{d}\right) \sin\left(\frac{\pi nz'}{d}\right) \\
&\times \frac{\exp[(ik_n-1/2L_n)|x-x'|]}{ik_nd}.
\end{split}
\end{equation}
Here $N_d=\llbracket kd/\pi\rrbracket$ is the total number of the propagating 
normal modes in the flat waveguide, which is determined by the integer part 
$\llbracket\ldots\rrbracket$ of the \emph{mode parameter} $kd/\pi$. Due to the 
quantization of the transverse wave number $k_{zn}=\pi n/d$, the quantum value 
of the longitudinal wave number is given by
\begin{equation}\label{eq:kn}
k_{xn}=k_{n}\equiv\sqrt{k^2-(\pi n/d)^2}, \qquad n=1, 2, ..., N_d.
\end{equation}
All other waveguide modes with $n>N_d$ are evanescent with imaginary values 
of $k_n$, and do not contribute to the transport properties. The average Green's 
function \eqref{eq:avGF} differs from the unperturbed one, in the appearance of 
the {\it mode-attenuation length} $L_n$.

%%%%%%%%%%%%%%%%%%%%%%%%%%%%%%%%%%%%%%%%%%%%%%%%%%%%%%%%%%%%%%%%%%%%%%%%%%%%%%%
\subsection{Mode-attenuation length}
\label{subsec:AttenuationLength}
%%%%%%%%%%%%%%%%%%%%%%%%%%%%%%%%%%%%%%%%%%%%%%%%%%%%%%%%%%%%%%%%%%%%%%%%%%%%%%%

The mode-attenuation length (or electron \emph{ total mean free path}) $L_n$ 
describes the scattering from the $n$th propagating mode into the others. From 
general theory of quasi-1D scattering systems it follows that the mode-attenuation 
length is determined by the \emph{backward} $L_n^{(b)}$, and \emph{forward} 
$L_n^{(f)}$ scattering lengths~\cite{BssFks_1979},
\begin{equation}\label{eq:Ln(f)(b)}
\frac{1}{L_n}=\frac{1}{L^{(b)}_{n}}+\frac{1}{L^{(f)}_{n}}.
\end{equation}
In accordance with Ref.~\onlinecite{RndIzrMkr-PRB-2007}, both backward and 
forward scattering lengths consist of two fundamentally different \emph{partial} 
lengths. One of these partial lengths is related to the \emph{amplitude} and 
\emph{gradient} mechanisms of surface scattering, whereas the other partial 
length is associated with the \emph{square-gradient} mechanism (the connection 
between these partial lengths and the scattering potential $\hat{U}(x,z)$ is 
discussed below). Therefore, backward and forward scattering lengths are given 
by
\begin{subequations}\label{eq:Ln(b)(f)}
\begin{eqnarray}
\label{eq:SRB_ASRB_Ln(b)}
\frac{1}{L_n^{(b)}}&=&\frac{1}{L^{(A,G)(b)}_{n}}+\frac{1}{L^{(SG)(b)}_{n}}, \\
\label{eq:SRB_ASRB_Ln(f)}
\frac{1}{L_n^{(f)}}&=&\frac{1}{L^{(A,G)(f)}_{n}}+\frac{1}{L^{(SG)(f)}_{n}}.
\end{eqnarray}
\end{subequations}
The partial lengths associated with the backscattering read
\begin{subequations}\label{eq:Ln(b)}
\begin{eqnarray}
\label{eq:Ln(G)(b)}
\frac{1}{L^{(A,G)(b)}_{n}}&=&\frac{\sigma^2}{d^6}
\sum_{n'=1}^{N_d}\; \frac{A_{nn'}}{k_n k_{n'}} W(k_n+k_{n'}), \\
\label{eq:Ln(SG)(b)}
\frac{1}{L^{(SG)(b)}_{n}}&=&\frac{\sigma^4}{d^4}
\sum_{n'=1}^{N_d}\; \frac{B_{nn'}}{k_n k_{n'}} \RSGP(k_n+k_{n'}).
\end{eqnarray}
\end{subequations}
The partial lengths associated with the forward scattering are similar to the 
backward ones but with the functions $W$ and $\RSGP$ depending on the 
argument $k_n-k_{n'}$, i.e.,
\begin{subequations} \label{eq:Ln(f)}
\begin{eqnarray}
\label{eq:Ln(G)(f)}
\frac{1}{L^{(A,G)(f)}_{n}}&=&\frac{\sigma^2}{d^6}
\sum_{n'=1}^{N_d}\; \frac{A_{nn'}}{k_n k_{n'}} W(k_n-k_{n'}), \\
\label{eq:Ln(SG)(f)}
\frac{1}{L^{(SG)(f)}_{n}}&=&\frac{\sigma^4}{d^4}
\sum_{n'=1}^{N_d}\; \frac{B_{nn'}}{k_n k_{n'}} \RSGP(k_n-k_{n'}).
\end{eqnarray}
\end{subequations}
Factors $A_{nn'}$ and $B_{nn'}$ depend upon the type of symmetry
exhibited by the opposite rough boundaries of the waveguide. Below, they are 
discussed and, in Table~\ref{tab:ABFactors_2Modes}, their values are explicitly 
shown for the SRB and ASRB two-mode waveguide.

The equations~\eqref{eq:Ln(b)} and \eqref{eq:Ln(f)} contain two different spectral 
functions. First, the \emph{roughness-height} power spectrum is included in 
Eqs.~\eqref{eq:Ln(G)(b)} and \eqref{eq:Ln(G)(f)}. This function is defined as the 
Fourier transform of the \emph{roughness-height} binary correlator
${\cal W}(x)$, see Eq.~\eqref{eq:<xixi>}),
\begin{equation}\label{eq:FT(W)}
W(k_x)=\int_{-\infty}^{\infty}dx\exp(-ik_xx)\,{\cal W}(x).
\end{equation}
Second, the formulas in Eqs.~\eqref{eq:Ln(SG)(b)} and \eqref{eq:Ln(SG)(f)} 
contain the so-called \emph{roughness-square-gradient} power spectrum,
\begin{equation}\label{eq:FT(W''2)}
\RSGP(k_x)=\int_{-\infty}^{\infty}dx\exp{(-ik_xx)}
{\cal W}''^2(x),
\end{equation}
where the double prime to the function ${\cal W}(x)$ denotes a second derivative 
with respect to $x$. Since ${\cal W}(x)$ and ${\cal W}''^2(x)$ are real and even 
functions of $x$, their Fourier transforms are even and real functions of the 
longitudinal wave number $k_x$. It should be also stressed that according to 
rigorous mathematical theorem, the power spectra are a non-negative functions 
of $k_x$ for any real random process $\xi(x)$.

The spectra \eqref{eq:FT(W)} and \eqref{eq:FT(W''2)} arise when deriving the 
correlator of the scattering potential $\hat{U}(x,z)$ in the $k_x$ representation. 
The correlator of $\hat{U}(x,z)$ emerges when averaging the Green's function 
within the perturvative approach~\cite{IzrMkrRnd-PRB_2006}. Specifically, the
terms in $\hat{U}(x,z)$ that depend upon the roughness \emph{amplitude} 
$\xi(x)$ and the roughness \emph{gradient} $\xi'(x)$ give rise to the terms in the 
correlator associated with $W(k_x)$, see the structure of $\hat{U}(x,z)$ in
Eqs.~\eqref{eq:SRB_U(x,z)} and \eqref{eq:ASRB_U(x,z)}. The terms that depend 
upon the roughness \emph{square gradient} $\xi'^2(x)$ lead to the terms in the 
correlator associated with the Fourier transform of 
$\langle \xi'(x)\xi'(x')\rangle^2={\cal W}''^2(x)$, which in fact is $\RSGP(k_x)$. 
It should be stressed that through the integration by parts the power spectrum of 
the roughness gradient $\xi'(x)$ can be reduced to $W(k_x)$. However, it is not 
possible to do the same for the power spectrum $\RSGP(k_x)$. This fact reflects 
a nontrivial role of the terms in the scattering potential containing $\xi'^2(x)$. 
Therefore these terms are associated with a specific and independent 
square-gradient surface-scattering mechanism.

Accordingly, the inverse scattering lengths given in \eqref{eq:Ln(b)(f)} are 
expressed as a sum of two terms that describe scattering governed by different 
surface scattering mechanisms. Thus, the \emph{amplitude-}scattering 
and \emph{gradient-}scattering mechanisms are associated with the terms 
$1/L_n^{(A,G)(b)}$ and $1/L_n^{(A,G)(f)}$ given in Eqs.~\eqref{eq:Ln(G)(b)} and 
\eqref{eq:Ln(G)(f)}, while the \emph{square-gradient-}scattering mechanism is 
associated with the terms $1/L_n^{(SG)(b)}$ and $1/L_n^{(SG)(f)}$ given by 
Eqs.~\eqref{eq:Ln(SG)(b)} and \eqref{eq:Ln(SG)(f)}).

It should be also stressed that, as functions of the correlation length $\Rc$, the 
mode-attenuation lengths $L_n^{(A,G)(b)}$ and $L_n^{(A,G)(f)}$ can behave 
very different from $L_n^{(SG)(b)}$ and $L_n^{(SG)(f)}$ because of their 
dependence upon the different power spectra $W(k_x)$ and $\RSGP(k_x)$. 
Owing to those peculiar behaviors, the competition between surface-scattering 
mechanisms emerges and, for some range of values of $\Rc$, the product 
$\sigma^4 S(k_x)$ can greatly increase its value in comparison with the quantity 
$\sigma^2 W(k_x)$, even for small $\sigma$. Thus, the lengths $L_n^{(SG)(b)}$ 
and $L_n^{(SG)(f)}$ can be compared, and even can be smaller than the lengths 
$L_n^{(A,G)(b)}$ and $L_n^{(A,G)(f)}$. A detailed discussion about this 
competition between surface-scattering mechanisms is presented in 
Section~\ref{sec:SRB}, for the SRB waveguide, and in Section~\ref{sec:ASRB}, 
for ASRB waveguide, with the specific power spectra 
\eqref{eq:GaussianPowerSpectra}, which are associated with the Gaussian 
correlator.

The approximation by which one arrives to Eqs.~\eqref{eq:avGF}, 
\eqref{eq:Ln(f)(b)} -- \eqref{eq:Ln(f)}, and the resulting domain of applicability of 
those equations were discussed in Ref.~\onlinecite{IzrMkrRnd-PRB_2006}. 
Specifically, it was shown that the validity of the results is restricted by
two independent criteria of weak surface scattering,
\begin{subequations}\label{eq:WeakSurfaceScattCriteria}
\begin{eqnarray}
\label{eq:WeakSurfaceScattCriteria1}
&\Lambda_n\ll2L_n,\\
\label{eq:WeakSurfaceScattCriteria2}
&\Rc\ll2L_n.
\end{eqnarray}
\end{subequations}
Here the cycle length, $\Lambda_n$, is the distance between two successive 
reflections of the $n$th mode from the rough boundaries,
\begin{equation}\label{eq:CycleLength}
\Lambda_n=k_nd/(\pi n/d).
\end{equation}
The criteria expressed in Eq.~\eqref{eq:WeakSurfaceScattCriteria} imply that the 
waves are weakly attenuated over both the correlation length $\Rc$ and cycle 
length $\Lambda_n$. Additionally, they restrict the corrugations to be small in 
height, $\sigma\ll d$. It should be also noted that statistical averaging is 
reasonable if the correlation length is much less than the waveguide length, 
$\Rc\ll L$. Therefore, the length $L$ must obey requirements similar to 
Eqs.~\eqref{eq:WeakSurfaceScattCriteria} formulated with respect to $L_n$. All 
these limitations are common in surface scattering theories that are based on an 
appropriate perturbative approach, see for example, 
Ref.~\onlinecite{BssFks_1979}.

%%%%%%%%%%%%%%%%%%%%%%%%%%%%%%%%%%%%%%%%%%%%%%%%%%%%%%%%%%%%%%%%%%%%%%%%%%%%%%%
\subsection{Reduction to two-mode waveguides}
\label{subsec:Two-modeWaveguides}
%%%%%%%%%%%%%%%%%%%%%%%%%%%%%%%%%%%%%%%%%%%%%%%%%%%%%%%%%%%%%%%%%%%%%%%%%%%%%%%

In what follows, this study is restricted to waveguides with two propagating modes,
\begin{equation}\label{eq:Nd=2}
N_d=\llbracket kd/\pi\rrbracket=2.
\end{equation}
In this case the mode parameter is confined within the interval,
\begin{equation}\label{eq:TwoModes_kRange}
2<kd/\pi<3,
\end{equation}
and the longitudinal wave numbers are given by
\begin{equation} \label{eq:k1k2}
k_{1}=\sqrt{k^2-(\pi/d)^2}, \quad k_{2}=\sqrt{k^2-(2\pi/d)^2}.
\end{equation}
Then, the factors $A_{nn'}$ and $B_{nn'}$, in Eqs.~\eqref{eq:Ln(b)} and 
\eqref{eq:Ln(f)}, which are drawn from Ref.~\onlinecite{RndIzrMkr-PRB-2007}, 
are displayed in Table~\ref{tab:ABFactors_2Modes}. The diagonal and 
off-diagonal elements are associated, respectively, with intra-mode scattering 
(scattering within the same mode) and inter-mode scattering (scattering from one 
mode into the other). Also, one can realize that the diagonal and off-diagonal 
elements of the matrices $A_{nn'}$ are related to the amplitude-scattering and 
gradient-scattering mechanisms, respectively.

\begin{table}
\caption{\label{tab:ABFactors_2Modes}
Matrices of factors $A_{nn'}$ and $B_{nn'}$ for the symmetric and antisymmetric 
two-mode surface-disordered waveguides.}
\begin{ruledtabular}
\begin{tabular}{rccc}
& & SRB & ASRB\\
\hline\\
\vspace{10pt}
$\left(\begin{array}{cc}
A_{11} & A_{12}\\
A_{21} & A_{22}
\end{array}\right)$ & $=$ &
$\left(\begin{array}{cc}
4\pi^4 & 0\\
0 & 64\pi^4
\end{array}\right)$,&
$\left(\begin{array}{cc}
0 & 16\pi^4\\
16\pi^4 & 0
\end{array}\right)$\\
\vspace{10pt}
$\left(\begin{array}{cc}
B_{11} & B_{12}\\
B_{21} & B_{22}
\end{array}\right)$ & $=$&
$\left(\begin{array}{cc}
\frac{(3+\pi^2)^2}{18} & 0\\
0 & \frac{(3+4\pi^2)^2}{18}
\end{array}\right)$,&
$\left(\begin{array}{cc}
\pi^4/2 & 0\\
0 & 8 \pi^4
\end{array}\right)$
\end{tabular}
\end{ruledtabular}
\end{table}

Bearing in mind this physical meaning of the entries $A_{nn'}$ and $B_{nn'}$, 
the structure of matrices in Table~\ref{tab:ABFactors_2Modes} reveals 
remarkable differences between the SRB and ASRB waveguides:
\begin{enumerate}
\item[(i)]
The attenuation in the SRB waveguide is determined by the competition between 
the amplitude-scattering and square-gradient-scattering mechanisms, which are 
both associated with the intra-mode scattering (in two-mode waveguide).
Therefore, the propagating modes in the SRB model should be regarded as 
independent, and, consequently, the SRB two-mode waveguide represents two 
independent 1D-surface-disordered channels. As is known, the transport 
properties of 1D disordered systems are described by the \emph{theory of 
Anderson localization} (see, e.g., Ref.~\onlinecite{IzrMkr_JPhysA_2005} and 
references therein). This theory has been already employed for the analysis of 
SRB single-mode waveguide \cite{MkrTrs_2001}. Now, the theory of 1D 
Anderson localization should be adapted for describing the transmittance of 
two-mode waveguide. According to this theory, the transport in each channel is 
completely specified by the scaling parameter that is the ratio between the 
sample length and corresponding backscattering length. Remarkably, the 
transmittance of any 1D disordered structure does not depend on the forward 
scattering.
This approach is presented in Section~\ref{sec:SRB}, where it is showed that 
$L/L_n^{(b)}$ becomes the scaling parameter of the transport through the $n$th 
channel.
\item[(ii)]
Otherwise, the attenuation in the ASRB waveguide is determined by the interplay 
between the gradient-scattering and square-gradient-scattering mechanisms. 
Since the former is related to the inter-mode scattering, the channels can become 
mixed and the theory that accounts for the transport properties of the ASRB 
waveguide should differ from that used in SRB waveguide. A general theory for 
the transmittance of waveguides that considers the interplay between both 
mechanisms is not presented here. However, the theoretical approach developed 
in Refs.~\onlinecite{IzrMkrRnd-PRB_2006,RndIzrMkr-PRB-2007} allows one to 
predict many of important characteristics of transport in different regimes. Here 
this approach is applied to the ASRB waveguide with two channels. As a result, 
it becomes possible to predict the interval of parameters in which only one of 
these mechanisms prevails. On the one hand, when the intra-mode scattering 
governs the attenuation, one deals with the same situation that was found in the 
SRB waveguide and, therefore, the transport properties may be specified by the 
scaling parameter that is the ratio between the sample length and corresponding 
backscattering length. On the other hand, when the inter-mode scattering 
determines the attenuation the transmittance of the $n$th propagating mode is 
specified by the scaling parameter $L/L_n$, which involves both backward 
$L_n^{(b)}$ and forward $L_n^{(f)}$ scattering lengths. The approach that 
considers intra-mode and inter-mode scattering in the ASRB waveguide is 
developed in Section~\ref{sec:ASRB}.
\end{enumerate}

The explicit expressions for the mode backward and forward scattering lengths, 
of our two-mode waveguides, are useful for continuous reference. From 
Eqs.~\eqref{eq:Ln(b)(f)} -- \eqref{eq:Ln(b)}, with the factors $A_{nn'}$ and 
$B_{nn'}$ from the column SRB in Table \ref{tab:ABFactors_2Modes}, the 
necessary backscattering lengths are given by
\begin{subequations}\label{eq:SRB_L1(b)(f)_L2(b)(f)}
\begin{eqnarray}
\label{eq:SRB_L1(b)}
\frac{1}{L_1^{(b)}}&\approx&4\pi^4\,\frac{\sigma^2}{d^6} 
\frac{W(2k_1)}{k_1^2}+9.2\,\frac{\sigma^4}{d^4} \frac{\RSGP(2k_1)}{k_1^2}, \\
\label{eq:SRB_L2(b)}
\frac{1}{L_2^{(b)}}&\approx&64\pi^4\,\frac{\sigma^2}{d^6} 
\frac{W(2k_2)}{k_2^2}+100.2\,\frac{\sigma^4}{d^4} \frac{\RSGP(2k_2)}{k_2^2}. \nonumber \\
\end{eqnarray}
\end{subequations}
Also from Eqs.~\eqref{eq:Ln(b)(f)} -- \eqref{eq:Ln(f)}, but with the factors $A_{nn'}$ 
and $B_{nn'}$ corresponding to the ASRB waveguide, the explicit expressions of 
the lengths read,
\begin{subequations}\label{eq:ASRB_L1(b)(f)_L2(b)(f)}
\begin{eqnarray}
\label{eq:ASRB_L1(b)}
\frac{1}{L_1^{(b)}}&=&16\pi^4\,\frac{\sigma^2}{d^6} \frac{W(k_1+k_2)}{k_1 k_2}
+\frac{\pi^4}{2}\frac{\sigma^4}{d^4} \frac{\RSGP(2k_1)}{k_1^2}, \nonumber \\ \\
\label{eq:ASRB_L1(f)}
\frac{1}{L_1^{(f)}}&=&16\pi^4\,\frac{\sigma^2}{d^6} \frac{W(k_1-k_2)}{k_1 k_2}
+\frac{\pi^4}{2}\frac{\sigma^4}{d^4} \frac{\RSGP(0)}{k_1^2}, \nonumber \\ \\
\label{eq:ASRB_L2(b)}
\frac{1}{L_2^{(b)}}&=&16\pi^4\,\frac{\sigma^2}{d^6} \frac{W(k_1+k_2)}{k_1 k_2}
+8\pi^4\,\frac{\sigma^4}{d^4} \frac{\RSGP(2k_2)}{k_2^2}, \nonumber \\ \\
\label{eq:ASRB_L2(f)}
\frac{1}{L_2^{(f)}}&=&16\pi^4\,\frac{\sigma^2}{d^6} \frac{W(k_1-k_2)}{k_1 k_2}
+8\pi^4\,\frac{\sigma^4}{d^4} \frac{\RSGP(0)}{k_2^2}. \nonumber \\
\end{eqnarray}
\end{subequations}

In this paper we consider the situation for which the correlator ${\cal W}(x)$ can 
be approximated by the Gaussian form,
${\cal W}(x)=\exp{(-x^2/2\Rc^2)}$. After the substitution of this correlator into 
formulas \eqref{eq:FT(W)} and \eqref{eq:FT(W''2)}, for any degree of roughness, 
the power spectra appearing in Eqs.~\eqref{eq:SRB_L1(b)(f)_L2(b)(f)} and 
\eqref{eq:ASRB_L1(b)(f)_L2(b)(f)} are given by
\begin{subequations}\label{eq:GaussianPowerSpectra}
\begin{eqnarray}
\label{eq:GaussianPowerSpectra_W}
W(k_x)&=&\sqrt{2\pi}\Rc\exp[-(k_x\Rc)^2/2],\\
\label{eq:GaussianPowerSpectra_S}
\RSGP(k_x)&=&\frac{\sqrt{\pi}}{16 \Rc^3}
[(k_x \Rc)^4-4(k_x \Rc)^2 + 12] \nonumber\\
&&\times\exp[-(k_x\Rc)^2/4].
\end{eqnarray}
\end{subequations}
These spectra have their maximum at $k_x=0$, but with $W(0)\sim\Rc$ and 
$\RSGP(0)\sim\Rc^{-3}$.

%%%%%%%%%%%%%%%%%%%%%%%%%%%%%%%%%%%%%%%%%%%%%%%%%%%%%%%%%%%%%%%%%%%%%%%%%%%%%%%
\section{Two-mode SRB Waveguide}
\label{sec:SRB}
%%%%%%%%%%%%%%%%%%%%%%%%%%%%%%%%%%%%%%%%%%%%%%%%%%%%%%%%%%%%%%%%%%%%%%%%%%%%%%%

%%%%%%%%%%%%%%%%%%%%%%%%%%%%%%%%%%%%%%%%%%%%%%%%%%%%%%%%%%%%%%%%%%%%%%%%%%%%%%%
\subsection{SRB: Competition between surface-scattering mechanisms}
\label{subsec:SRBCompetition}
%%%%%%%%%%%%%%%%%%%%%%%%%%%%%%%%%%%%%%%%%%%%%%%%%%%%%%%%%%%%%%%%%%%%%%%%%%%%%%%

In order to discuss the competition between various mechanisms of surface 
scattering, we should distinguish between two regions for the correlation length 
$\Rc$. These regions are denoted as the region of \emph{small-scale roughness} 
($k\Rc\ll1$) and the region of \emph{large-scale roughness} ($k\Rc\gg1$). Both 
regions correspond to weak correlations between successive reflections of the 
wave from rough boundaries ($\Rc\ll\Lambda_n$). Note that due to the last 
requirement the second of the weak-scattering conditions 
\eqref{eq:WeakSurfaceScattCriteria} is satisfied automatically when the first one 
is met, $\Rc\ll\Lambda_n\ll2L_n,L$.

%%%%%%% R1.
\emph{\bf Small-scale-roughness - } In this region the surface roughness can be 
regarded as a delta-correlated random process of the white-noise type. Taking 
into account the evident relationship $1\lesssim k\Lambda_n$ and the 
weak-scattering conditions \eqref{eq:WeakSurfaceScattCriteria}, one can get the 
following inequalities to specify this region together with requirements of 
applicability of the theory:
\begin{equation}\label{eq:region1_SSR}
k\Rc\ll 1\lesssim k\Lambda_n\ll2kL_n.
\end{equation}
Under the condition \eqref{eq:region1_SSR}, the argument of the power spectra 
\eqref{eq:GaussianPowerSpectra} is much less than the scale of their decrease, 
$k_x\ll R^{-1}$. Therefore, one can write
\begin{subequations} \label{eq:Gaussian_SSR}
\begin{eqnarray}
\label{eq:GaussianW_SSR}
W(k_x)&\approx& W(0)=\sqrt{2 \pi} \Rc,\\
\label{eq:GaussianT_SSR}
\RSGP(k_x)&\approx& \RSGP(0)=3\sqrt{\pi}/4\Rc^3.
\end{eqnarray}
\end{subequations}

In this case, when $\Rc$ decreases, the spectrum $\RSGP(k_x)$ increases as 
fast as $\Rc^{-3}$, whereas $W(k_x)$ decreases as $\Rc$. As one can see from 
Eqs.~\eqref{eq:SRB_L1(b)(f)_L2(b)(f)}, in spite of the fact that the first term is 
proportional to $\sigma^2$ while the second is proportional to $\sigma^4$, for 
any value of the roughness height $\sigma$, there is a region of small values of 
$\Rc$, where the square-gradient-scattering mechanism predominates (provided 
the condition \eqref{eq:region1_SSR} is fulfilled). Specifically, if the correlation 
length $\Rc$ is smaller than some \emph{crossing point} $\Rc_{\cpt n}$, the first 
term in Eqs.~\eqref{eq:SRB_L1(b)} and \eqref{eq:SRB_L2(b)} can be neglected 
and the mode-backscattering lengths are approximated to
\begin{subequations}\label{eq:SSRApprox1_Ln(b)}
\begin{eqnarray}
\label{eq:SSRApprox1_L1(b)}
\frac{1}{L_{1}^{(b)}}&\approx&12.2\;k_1\,
\frac{(\sigma/d)^4}{(k_1 R)^3}
\quad\text{for}\;k\Rc\ll k\Rc_{\cpt1}\ll1,\nonumber\\ \\
\label{eq:SSRApprox1_L2(b)}
\frac{1}{L_{2}^{(b)}}&\approx&133.3\;k_2\,
\frac{(\sigma/d)^4}{(k_2 R)^3}
\quad\text{for}\;k\Rc\ll k\Rc_{\cpt2}\ll1.\nonumber\\
\end{eqnarray}
\end{subequations}
Otherwise, when $\Rc$ is larger than $\Rc_{\cpt n}$, the square-gradient 
(second) term in Eqs.~\eqref{eq:SRB_L1(b)} and \eqref{eq:SRB_L2(b)} can now 
be neglected and the mode-backscattering lengths are approximated to
\begin{subequations}\label{eq:SSRApprox2_Ln(b)}
\begin{eqnarray}
\label{eq:SSRApprox2_L1(b)}
\frac{1}{L_{1}^{(b)}}&\approx&10.0\;k_1\,
\frac{(\sigma/d)^2\,(k_1 R)}{(k_1 d/\pi)^4}\nonumber\\
&&\qquad\qquad \text{for}\;k\Rc_{\cpt1}\ll k\Rc\ll1, \\
\label{eq:SSRApprox2_L2(b)}
\frac{1}{L_{2}^{(b)}}&\approx&160.4\;k_2\,
\frac{(\sigma/d)^2\,(k_2 R)}{(k_2 d/\pi)^4}\nonumber\\
&&\qquad\qquad\text{for}\;k\Rc_{\cpt2}\ll k\Rc\ll1.
\end{eqnarray}
\end{subequations}

The dimensionless crossing point $k\Rc_{\cpt n}$ of the $n$th backscattering 
length can be located either on the border between the regions of small- and 
large-scale-roughness, or inside the first region. In the former case, 
$k\Rc_{\cpt n}\sim1$, and within whole region \eqref{eq:region1_SSR}, the 
mode-backscattering lengths are contributed mainly by 
square-gradient-scattering, i.e. by Eq.~\eqref{eq:SSRApprox1_Ln(b)}. In the 
latter case, $k\Rc_{\cpt1}$ and $k\Rc_{\cpt2}$ are found by searching for the 
intersection of the asymptote \eqref{eq:SSRApprox1_L1(b)} with 
\eqref{eq:SSRApprox2_L1(b)}, and \eqref{eq:SSRApprox1_L2(b)} with 
\eqref{eq:SSRApprox2_L2(b)}. Thus, the crossing points read
\begin{equation}
k\Rc_{\cpt 1}\approx k\Rc_{\cpt 2}\approx\;0.3\;kd\; \sqrt{\sigma/d}\;.
\label{eq:SRB_kR_c_n}
\end{equation}

%%%%%%% R2.
\emph{\bf Large-scale-roughness - }
This region arises when the correlation length $\Rc$ becomes much larger than 
the wave length $2\pi/k$, but still remains much less that the cycle length 
$\Lambda_n$,
\begin{equation}\label{eq:region2_LSR_WC}
1\ll k\Rc\ll k\Lambda_n\ll2kL_n.
\end{equation}
Here the square-gradient term in Eqs.~\eqref{eq:SRB_L1(b)} and 
\eqref{eq:SRB_L2(b)} can be neglected and the mode-backscattering lengths 
are approximated to
\begin{subequations}\label{eq:LSRWCApprox_Ln(b)}
\begin{eqnarray}
\label{eq:LSRWCApprox_L1(b)}
\frac{1}{L_{1}^{(b)}}&\approx&10.0\;k_1\,
\frac{(\sigma/d)^2\,(k_1R)}{(k_1d/\pi)^4}\;\exp\left[-2(k_1R)^2\right]
\nonumber\\
&&\qquad\qquad\text{for}\;1\ll k\Rc\ll k\Lambda_1,\\
\label{eq:LSRWCApprox_L2(b)}
\frac{1}{L_{2}^{(b)}}&\approx&160.4\;k_2\,
\frac{(\sigma/d)^2\,(k_2R)}{(k_2d/\pi)^4}\;\exp\left[-2(k_2R)^2\right]
\nonumber\\
&&\qquad\qquad\text{for}\;1\ll k\Rc\ll k\Lambda_2.
\end{eqnarray}
\end{subequations}

%%%%%%%%%%%%%%%%%%%%%%%%%%%%%%%%%%%%%%%%%%%%%%%%%%%%%%%%%%%%%%%%%%%%%%%%%%%%%%%
\subsection{SRB: Total and mode transmittances}
%%%%%%%%%%%%%%%%%%%%%%%%%%%%%%%%%%%%%%%%%%%%%%%%%%%%%%%%%%%%%%%%%%%%%%%%%%%%%%%

Based on the diagonal form of matrices $A_{nn'}$ and $B_{nn'}$ of the SRB 
waveguide, this study argues that the attenuation of waves arises due to the 
intra-mode scattering only (see Table.~\ref{tab:ABFactors_2Modes}). 
Consequently, the two-mode waveguide can be treated as two independent 
1D-surface-disordered wires with their corresponding backscattering lengths 
$L_1^{(b)}$ and $L_2^{(b)}$. Thus, the known theory of 1D localization becomes 
pertinent here. In adapting this theory to the present situation, it should be 
realized that the scaling parameters $L/L_1^{(b)}$ and $L/L_2^{(b)}$ determine, 
respectively, the average transmittance of the first and second mode. Specifically, 
the transmittance of the $n$th mode ($n=1,2$) can be taken, for example, from 
Ref.~\onlinecite{IzrMkr_JPhysA_2005}. It reads
\begin{equation}\label{eq:SRB-TM-def}
\begin{split}
\langle T_M(L/L_n^{(b)})\rangle=&\frac{1}{2\sqrt{\pi}}
\left(\frac{L}{4L_n^{(b)}}\right)^{-3/2}
\exp\left(-\frac{L}{4L_n^{(b)}}\right)\\
&\times\int_0^\infty \frac{z^2\, dz}{\cosh z}
\exp\left(-z^2\frac{L_n^{(b)}}{L}\right).
\end{split}
\end{equation}
The total transmittance is given by,
\begin{equation}\label{eq:SRB-Twn}
\langle T(L)\rangle=\sum_{n=1}^{2} \langle T_M(L/L_n^{(b)})\rangle.
\end{equation}
Eqs.~ \eqref{eq:SRB-TM-def} and \eqref{eq:SRB-Twn} can be directly found from 
the general expression \eqref{eq:T(L)-def} with the use of well developed 
methods, such as, e.g., the perturbative diagrammatic technique of Berezinski 
\cite{Brz_JETP_1973,AbrRyz_AdvPhys_1978}, the invariant imbedding method 
\cite{BllWng_1975,Klt_1986} or the two-scale approach
\cite{MkrTrs_1998,Mkr_TR_1999,MkrTrs_2001}. Here it should be stressed that 
the ratio $L/L_n^{(b)}$ includes the contribution of both the amplitude-scattering 
mechanism and the square-gradient-scattering mechanism through the first and 
second terms in Eqs.~\eqref{eq:SRB_L1(b)} and \eqref{eq:SRB_L2(b)}.

\begin{figure}
\includegraphics[width=8cm]{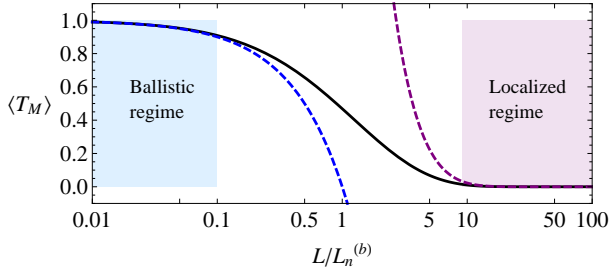}
\caption{\label{fig:SRBTMandAsymptotes}
(Color online) SRB waveguide: the mode transmittance \eqref{eq:SRB-TM-def}
(solid line) and its asymptotes \eqref{eq:SRB-TM-bal} and
\eqref{eq:SRB-TM-loc} for ballistic and localized regimes, respectively (dashed 
lines).}
\end{figure}

The mode transmittance \eqref{eq:SRB-TM-def} exhibits the ballistic behavior for 
large backscattering length,
\begin{equation}\label{eq:SRB-TM-bal}
\langle T_M(L/L_n^{(b)})\rangle \approx 1-L/L_n^{(b)}
\quad\text{for}\quad L\ll L_n^{(b)}.
\end{equation}
In this case the $n$th conducting channel is practically transparent. On the 
contrary, the the mode transmittance displays exponential decrease as the 
waveguide length $L$ exceeds $4L_n^{(b)}$,
\begin{multline}\label{eq:SRB-TM-loc}
\langle T_M(L/L_n^{(b)})\rangle\approx\frac{\pi^{5/2}}{16}
\left(\frac{L}{4L_n^{(b)}}\right)^{-3/2}
\exp\left(-\frac{L}{4L_n^{(b)}}\right) \\
\text{for}\quad L_n^{(b)}\ll L.
\end{multline}
The dependence of the function \eqref{eq:SRB-TM-def} and its asymptotes 
\eqref{eq:SRB-TM-bal} and \eqref{eq:SRB-TM-loc} on the scaling parameter 
$L/L_n^{(b)}$ are shown in Fig.~\ref{fig:SRBTMandAsymptotes}.

In accordance with two mode-transport regimes \eqref{eq:SRB-TM-bal}, 
\eqref{eq:SRB-TM-loc} and due to the hierarchy of the mode-backscattering 
lengths, $L_2^{(b)}<L_1^{(b)}$ (this hierarchy can be seen when comparing 
Eq.~\eqref{eq:SRB_L1(b)} with \eqref{eq:SRB_L2(b)}, or more directly through 
the couples of Eqs. \eqref{eq:SSRApprox1_Ln(b)}, \eqref{eq:SSRApprox2_Ln(b)} 
and \eqref{eq:LSRWCApprox_Ln(b)}), the total transmittance \eqref{eq:SRB-Twn} 
has the following three transport regimes:
\begin{enumerate}
\item[(i)]
In the \emph{regime of localization} when the largest mode-backscattering length 
$L_1^{(b)}$ is smaller than the waveguide length $L$,
\begin{equation}\label{eq:SRB_LocRegime}
1<L/L_1^{(b)}<L/L_2^{(b)},
\end{equation}
both propagating modes are strongly localized, and their transmittances are 
exponentially small. The total transmittance \eqref{eq:SRB-Twn} is 
approximately equal to the transmittance of the first mode obeying the asymptote 
\eqref{eq:SRB-TM-loc},
\begin{equation}\label{eq:SRB-T-loc}
\begin{split}
\langle T(L)\rangle&\approx \langle T_M(L/L_1^{(b)})\rangle \\
&\approx \frac{\pi^{5/2}}{16}\left(\frac{L}{4L_1^{(b)}}\right)^{-3/2}
\exp\left(-\frac{L}{4L_1^{(b)}}\right).
\end{split}
\end{equation}
The waveguide is non-transparent in this regime.
\item[(ii)]
The \emph{coexistence regime} arises when the smallest backscattering length 
$L_2^{(b)}$ is smaller, while the largest backscattering length $L_1^{(b)}$ is 
larger than the waveguide length $L$,
\begin{equation}\label{eq:SRB_IntRegime}
L/L_1^{(b)}<1<L/L_2^{(b)}.
\end{equation}
In this case, the first mode manifests the ballistic behavior \eqref{eq:SRB-TM-bal}, 
while the second mode is still localized in line with Eq.~\eqref{eq:SRB-TM-loc}. 
Therefore, as before, the total transmittance \eqref{eq:SRB-Twn} is determined 
by the transmittance of the first mode, however governed by the ballistic 
asymptote \eqref{eq:SRB-TM-bal},
\begin{equation}\label{eq:SRB-T-Int}
\langle T(L)\rangle\approx\langle T_M(L/L_1^{(b)})\rangle\approx1.
\end{equation}
\item[(iii)]
The \emph{ballistic regime} emerges under the conditions
\begin{equation}\label{eq:SRB_BalRegime}
L/L_1^{(b)}< L/L_2^{(b)}< 1,
\end{equation}
when the smallest mode-backscattering length $L_2^{(b)}$ is larger than $L$. As 
a consequence, both conducting channels are open having almost unit 
transmittances \eqref{eq:SRB-TM-bal}. The waveguide is almost perfectly 
transparent. Its total transmittance \eqref{eq:SRB-Twn} approximately equals to 
the total number of the propagating modes,
\begin{equation}\label{eq:SRB-T-Bal}
\langle T(L)\rangle\approx2.
\end{equation}
\end{enumerate}

%%%%%%%%%%%%%%%%%%%%%%%%%%%%%%%%%%%%%%%%%%%%%%%%%%%%%%%%%%%%%%%%%%%%%%%%%%%%%%%
\subsection{SRB: Transmittance vs correlation length}
%%%%%%%%%%%%%%%%%%%%%%%%%%%%%%%%%%%%%%%%%%%%%%%%%%%%%%%%%%%%%%%%%%%%%%%%%%%%%%%

\begin{figure}[ht]
\includegraphics[width=\columnwidth]{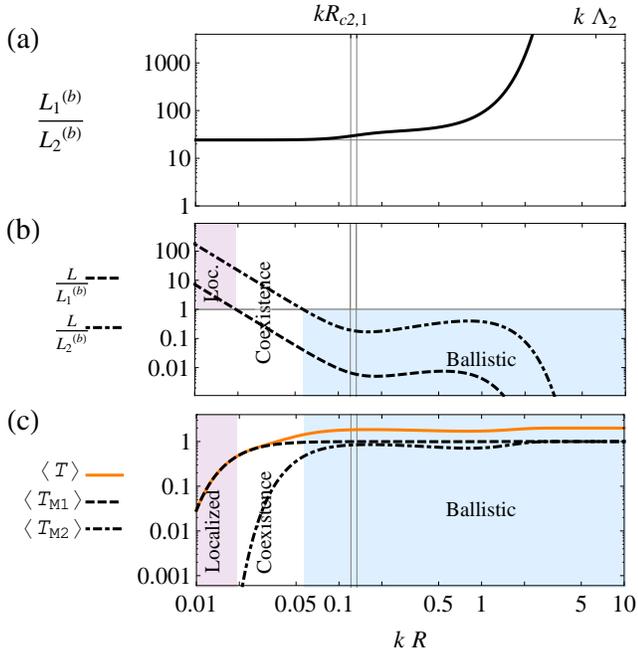}
\caption{\label{fig:SRBarrayVSkR}
(Color online) SRB waveguide: Mode-backscattering lengths, $L_n^{(b)}$, and 
transmittances, $\langle T_{Mn}\rangle=\langle T_M(L/L_n^{(b)})\rangle$, 
$\langle T(L)\rangle$, vs dimensionless correlation length, $k\Rc$, for 
$\sigma/d=0.0025$, $kd/\pi=2.54$, ($k\sigma\approx 0.02$), and $L/d=1600$. 
The crossing points $k\Rc_{c1}\approx 0.13$ and $k\Rc_{c2}\approx 0.12$ are 
indicated on the top of the upper plot. The boundary between the localization and 
the coexistence regime is at $k\Rc\approx 0.02$ and with further increment of 
$k\Rc$ the propagation regime becomes ballistic after the point 
$k\Rc\approx 0.06$.}
\end{figure}

Here we discuss the dependence of the mode-backscattering lengths and 
transmittances on the dimensionless correlation parameter $k\Rc$. For this, we 
plot the ratio $L_1^{(b)}/L_2^{(b)}$, as well as the scaling parameters 
$L/L_1^{(b)}$ and $L/L_2^{(b)}$, see Figs.~\ref{fig:SRBarrayVSkR} (a) and (b). 
These plots are computed from the expressions for $L_n^{(b)}$ given in 
Eqs.~\eqref{eq:SRB_L1(b)} and \eqref{eq:SRB_L2(b)} with power spectra 
\eqref{eq:GaussianPowerSpectra}. If $\Rc$ decreases, the lengths $L_n$ 
decreases as $\Rc^3$ whereas $\Lambda_n$ remains fixed, thus, the smallest 
value of $\Rc$ is restricted by the criterion \eqref{eq:WeakSurfaceScattCriteria1}; 
this criterion is fulfilled since $\Lambda_1/2L_1\approx 0.01 \ll 1$ and
$\Lambda_2/2L_2\approx 0.04 \ll 1$ at the left boundary of the plots, 
$k\Rc=0.01$. Because the plots illustrate the case of weak correlations 
($\Rc\ll \Lambda_n$), the condition 
\eqref{eq:WeakSurfaceScattCriteria2} is satisfied automatically if
\eqref{eq:WeakSurfaceScattCriteria1} is met.

Figure~\ref{fig:SRBarrayVSkR} (a) exhibits the crossover from the 
square-gradient to amplitude scattering. Specifically, with an increase of $k\Rc$ 
the ratio $L_1^{(b)}/L_2^{(b)}$ starts with the value,
\begin{equation}\label{eq:AprxR1_L1_over_L2_SRB}
\frac{L_1^{(b)}}{L_2^{(b)}}\approx10.9\left(\frac{k_1}{k_2}\right)^2
\qquad\text{for}\quad k\Rc\ll k\Rc_{\cpt2},
\end{equation}
and grows in accordance with the relation,
\begin{multline}\label{eq:AprxR2_L1_over_L2_SRB}
\frac{L_1^{(b)}}{L_2^{(b)}}\approx
16\left(\frac{k_1}{k_2}\right)^2\exp\left[6\pi^2 (R/d)^2\right]\\
\text{for}\quad k\Rc_{\cpt1}\ll k\Rc\ll k\Lambda_2.
\end{multline}
The asymptotic behaviors \eqref{eq:AprxR1_L1_over_L2_SRB} and
\eqref{eq:AprxR2_L1_over_L2_SRB} are obtained, respectively, from 
Eqs.~\eqref{eq:SSRApprox1_Ln(b)} and \eqref{eq:LSRWCApprox_Ln(b)}. Note 
that the hierarchy $L_1^{(b)}>L_2^{(b)}$ remains for any value of the parameters. 
With the curves in Fig.~\ref{fig:SRBarrayVSkR} (b), the different transport regimes 
defined in Eqs.~\eqref{eq:SRB_LocRegime} -- \eqref{eq:SRB-T-Bal} can be 
identified.

The mode transmittances \eqref{eq:SRB-TM-def} and total transmittance 
\eqref{eq:SRB-Twn} are plotted in Fig.~\ref{fig:SRBarrayVSkR} (c). Because of 
the parameters used in the plots, the boundary between the localization and the 
coexistence regime is at $k\Rc\approx 0.02$. With further increment of $k\Rc$ 
the propagation regime becomes ballistic after the point $k\Rc\approx 0.06$. 
Figure~\ref{fig:SRBarrayVSkR} (c) clearly demonstrates that, due to the hierarchy 
of backscattering lengths, $L_2^{(b)}<L_1^{(b)}$, the transmittance of the second 
mode is always smaller than the transmittance of the first one,
\begin{equation}\label{eq:SRB-T2<T1}
\langle T_M(L/L_2^{(b)})\rangle<\langle T_M(L/L_1^{(b)})\rangle.
\end{equation}
Therefore, within the localization \eqref{eq:SRB_LocRegime} and coexistence 
\eqref{eq:SRB_IntRegime} regions the total transmittance \eqref{eq:SRB-Twn} is 
mainly contributed by the first term, i.e. by the first-mode transmittance,
\begin{equation}\label{eq:SRB-T=T1}
\langle T(L)\rangle\approx\langle T_M(L/L_1^{(b)})\rangle.
\end{equation}
Within the localization regime, $\langle T(L)\rangle$ behaves as indicated in 
Eq.~\eqref{eq:SRB-T-loc}. Within the coexistence regime the total transmittance 
is described by Eq.~\eqref{eq:SRB-T-Int}. For the ballistic regime 
\eqref{eq:SRB_BalRegime}, the approximation given by 
Eq.~\eqref{eq:SRB-TM-bal} describes well the behavior of the mode 
transmittances. Correspondingly, the value of the total transmittance tends to the 
total number $N_d=2$ of the propagating modes, see Eq.~\eqref{eq:SRB-T-Bal}.

Thus, with the variation of roughness correlations one can realize all three 
regimes, which is inherent for the transport through the two-mode SRB 
waveguide. Note one more that coexistence regime \eqref{eq:SRB_IntRegime}, 
\eqref{eq:SRB-T-Int} arises only due to the fact that in the SRB waveguide the 
hierarchy of the mode-backscattering lengths is always present.

%%%%%%%%%%%%%%%%%%%%%%%%%%%%%%%%%%%%%%%%%%%%%%%%%%%%%%%%%%%%%%%%%%%%%
\subsection{SRB: Transmittance vs mode parameter}
%%%%%%%%%%%%%%%%%%%%%%%%%%%%%%%%%%%%%%%%%%%%%%%%%%%%%%%%%%%%%%%%%%%%%

\begin{figure}
\includegraphics[width=\columnwidth]{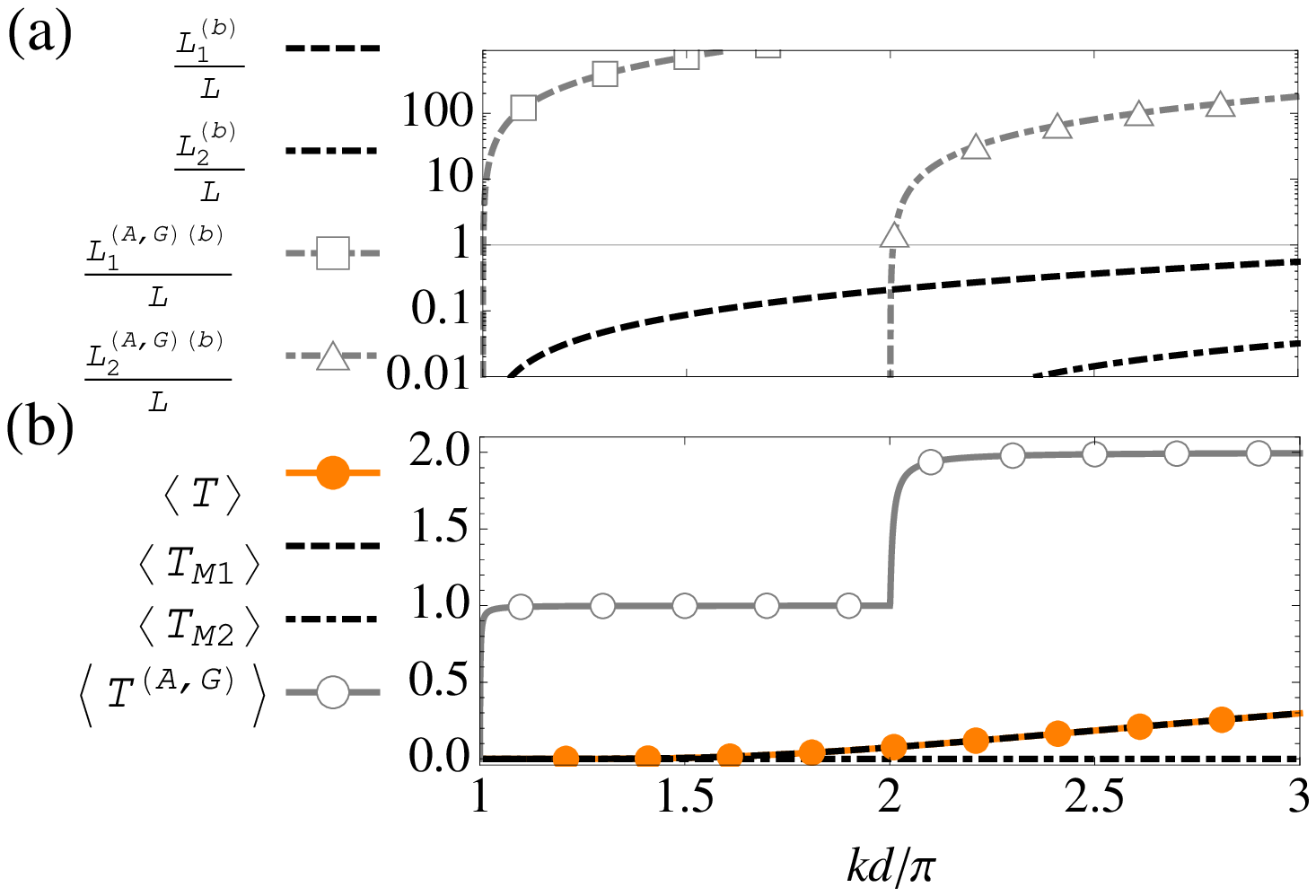} \\
\includegraphics[width=\columnwidth]{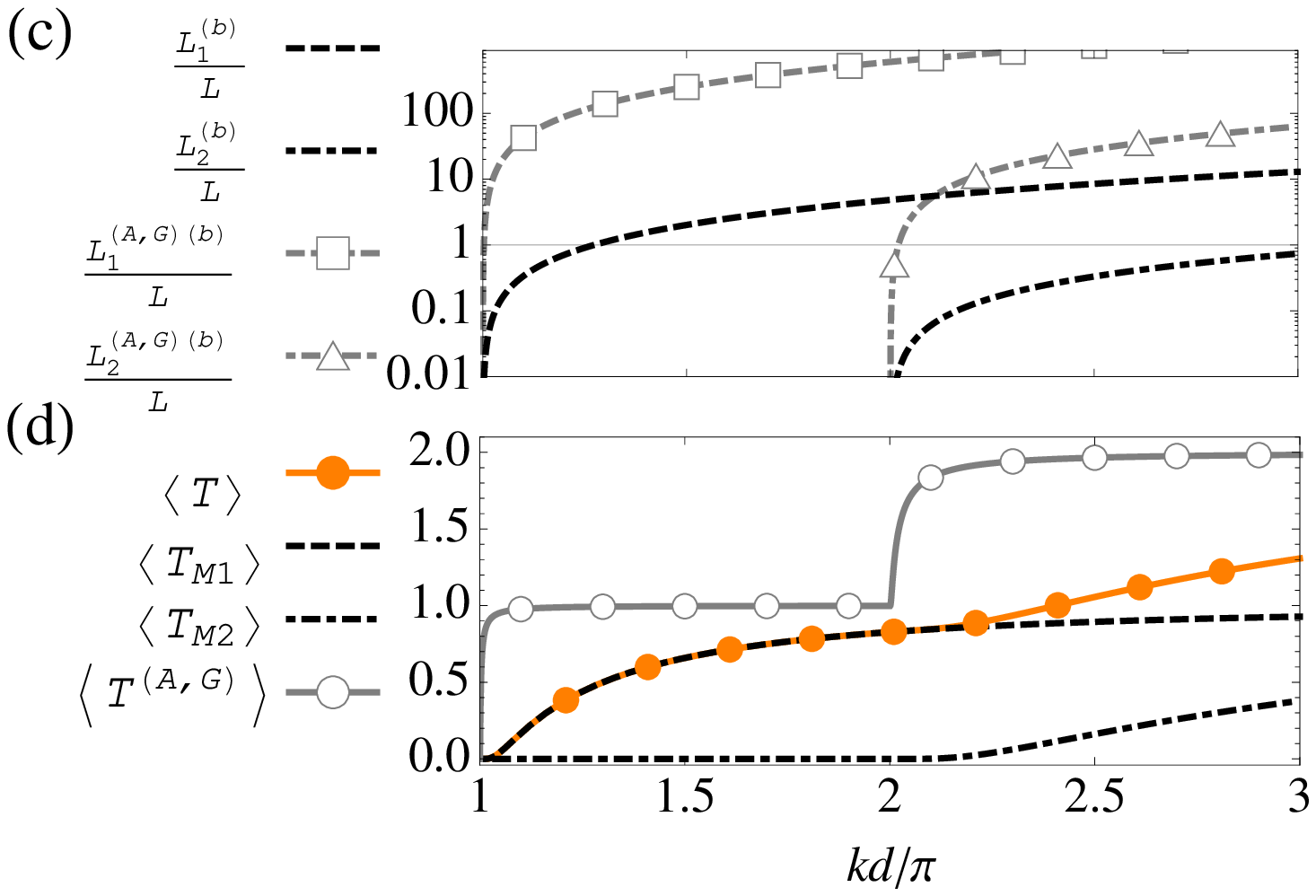} \\
\includegraphics[width=\columnwidth]{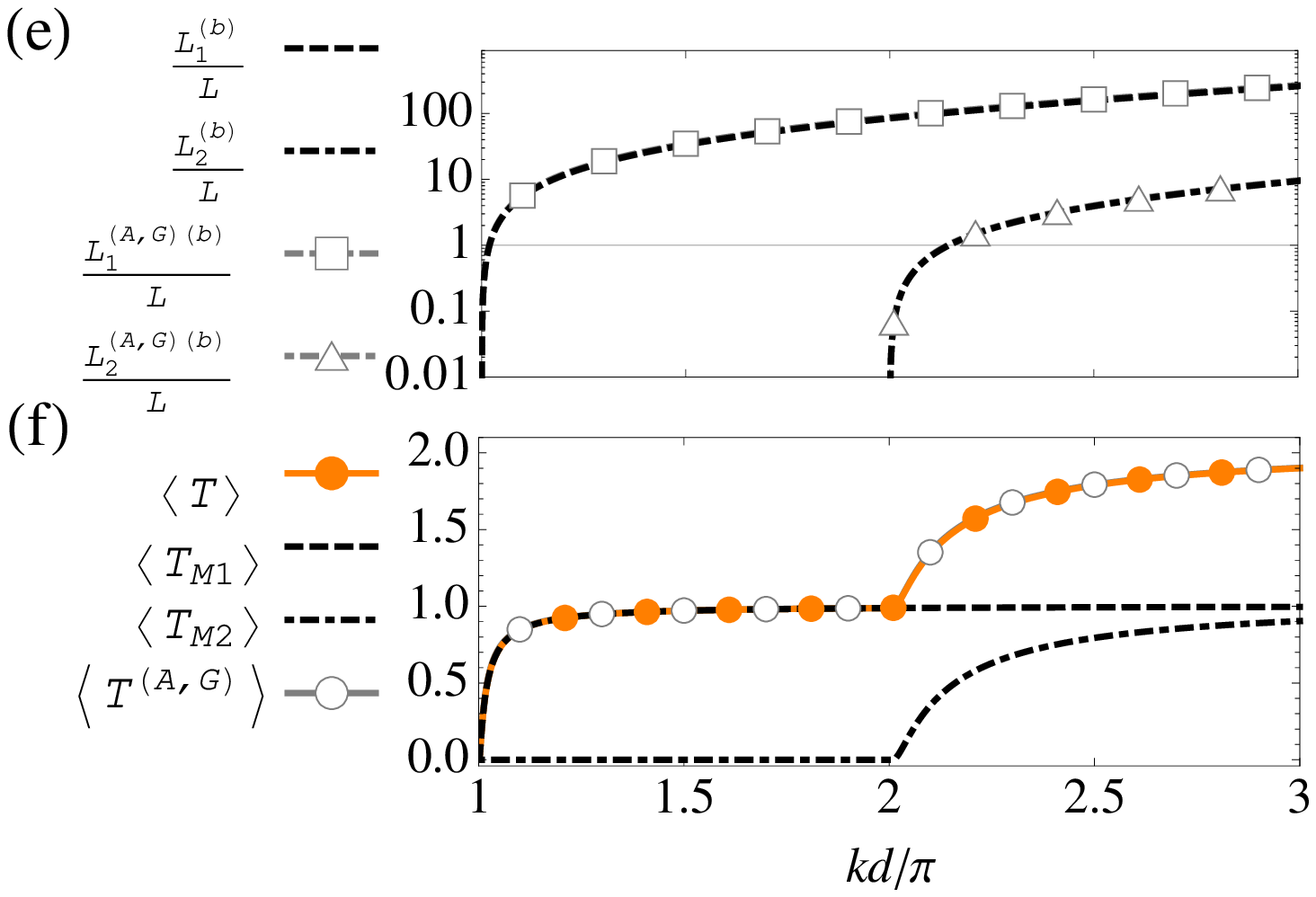}
\caption{\label{fig:SRBarrayVSkdOverPi}
(Color online) SRB waveguide: Mode-backscattering lengths and transmittances
versus mode parameter for $\sigma/d=0.0025$, $L/d=1600$, with
$\Rc/d\approx 0.0018$ for (a)-(b), $\Rc/d\approx 0.005$ for (c)-(d) and 
$\Rc/d\approx 0.038$ for (e)-(f). Note that for fixed values of the roughness height 
$\sigma$ and the mode parameter within the interval $2<kd/\pi<3$, $k\sigma$ 
ranges the interval $0.016\lessapprox k\sigma \lessapprox 0.024$; similarly, for 
fixed values of the correlation length $\Rc$ and the mode parameter within the 
interval $2<kd/\pi<3$, the dimensionless correlation length $k\Rc$ ranges the 
interval
$0.011\lessapprox k\Rc\lessapprox 0.017$ in (a)-(b),
$0.031\lessapprox k\Rc\lessapprox 0.047$ in (c)-(d) and
$0.23\lessapprox k\Rc\lessapprox 0.35$ in (e)-(f).}
\end{figure}

From the experimental viewpoint, it could be more feasible to study the transport 
properties of a waveguide by fixing its geometrical parameters such as length, 
width, etc., whereas the wave number $k$ is varied within some interval. In 
Fig.~\ref{fig:SRBarrayVSkdOverPi} we present three pairs of frames formed by 
plots of mode-backscattering lengths and transmittances as functions of the mode 
parameter $kd/\pi$ (dimensionless wave number) within the range of first two 
modes. The pairs (a)-(b), (c)-(d) and (e)-(f) mainly illustrate, respectively, the 
localized \eqref{eq:SRB_LocRegime}, coexistence \eqref{eq:SRB_IntRegime}
and ballistic \eqref{eq:SRB_BalRegime} transport regimes. Here the different 
regimes are illustrated by selecting three specific values of the correlation length 
$\Rc$ whereas the length of the waveguide $L$ stays constant; when 
$kd/\pi=2.54$, the mode-backscattering lengths and transmittances corresponds 
to those presented in Fig.~\ref{fig:SRBarrayVSkR} at the points $k\Rc=0.014$, 
$k\Rc=0.04$ and $k\Rc=0.3$. In this way, inside each regime, the relative 
influence of the scattering mechanisms is different. Another possibility to arrive to 
different transport regimes, which is not shown here, would consist in selecting
only one value for $k\Rc$ and three different values of the length $L$.

Figure~\ref{fig:SRBarrayVSkdOverPi} (a), (c) and (e) shows the plots of 
$L_n^{(b)}/L$ computed from Eqs.~\eqref{eq:SRB_L1(b)} and 
\eqref{eq:SRB_L2(b)}, as well as $L_n^{(A,G)(b)}/L$, which also comes from 
these equations but when the square-gradient term is neglected. If $kd/\pi$ 
increases, the $n$th attenuation length also increases from its value equal to 
zero at $kd/\pi=n$, where the $n$th normal mode opens ($n=1,2$). Note that our 
results, which are obtained within the second order approximation in the 
perturbation potential, can only indicate the position of these zeros and cannot 
correctly describe their line shape in the narrow vicinity of integer values of 
$kd/\pi$. The reason is that at these resonance points the longitudinal wave 
number $k_n$ vanishes and, as a consequence, the corresponding inverse 
attenuation length diverges. Conventionally, this problem is solved within the 
self-consistent approach providing finite resonance values for $1/L_n^{(b)}$
(see, e.g., Ref.~\onlinecite{BssFks_1979}). It should be mentioned that the curves 
associated with the first propagating mode do not manifest any resonance at 
$kd/\pi=2$, even with the opening of the second propagating mode. This occurs 
since both mechanisms that govern the scattering in two-mode SRB waveguide 
are associated with intra-mode transitions only. Remarkably, in frames (a) and (c) 
the dimensionless lengths $L^{(A,G)(b)}_1/L$ and $L^{(A,G)(b)}_2/L$ are much 
larger than the total lengths, $L_1^{(b)}/L$ and $L_2^{(b)}/L$. This is because the 
main contribution is due to the square-gradient-scattering lengths, 
$L^{(SG)(b)}_1/L$ and $L^{(SG)(b)}_2/L$. Therefore, for values of the mode 
parameter far from the resonances, the approximate behavior of $L_n^{(b)}/L$ is 
given by Eqs.~\eqref{eq:SSRApprox1_Ln(b)} (not plotted in Figure). Otherwise, 
in frame (e), in which the main contribution is due to the 
amplitude-gradient-scattering lengths, $L^{(A,G)(b)}_1/L$ and 
$L^{(A,G)(b)}_2/L$, the approximate behavior of $L_n^{(b)}/L$ is given by 
Eqs.~\eqref{eq:SSRApprox2_Ln(b)}.

The mode transmittances \eqref{eq:SRB-TM-def}, and total transmittance 
\eqref{eq:SRB-Twn} are plotted in Fig.~\ref{fig:SRBarrayVSkdOverPi} (b), (d) and 
(f). In frame (b) the first and second propagating modes are in localized transport 
regime; this fact is indicated by the small mode transmittances given by 
Eq.~\eqref{eq:SRB-TM-loc}. The total transmittance is approximately equal to the 
transmittance of the first mode, thus, it is given by Eq.~\eqref{eq:SRB-T-loc}. As 
one can see in Fig.~\ref{fig:SRBarrayVSkR}, with the increment of $k\Rc$ the 
mode-attenuation lengths also increases resulting in the condition 
\eqref{eq:SRB_IntRegime} that is illustrated in frame (d). Here, the first and 
second propagating modes are, respectively, in ballistic and localized transport 
regimes. The coexistence situation for the total transmittance is graphically 
indicated by a very small deviation of $\langle T(L)\rangle$ from 
$\langle T_M(L/L_1^{(b)})\rangle$, when $kd/\pi$ approaches 3. This behavior is 
marked by the ballistic asymptote \eqref{eq:SRB-TM-bal} for 
$\langle T_M(L/L_1^{(b)})\rangle$ and by increasing rate of the curve 
$\langle T_M(L/L_2^{(b)})\rangle$ with the localization asymptote 
\eqref{eq:SRB-TM-loc}. The predominating contribution of the 
square-gradient-scattering mechanism can be noted visually in frame (b) and (d) 
by comparing the curve of $\langle T(L)\rangle$ with the curve 
$\langle T^{(A,G)}(L)\rangle$,
\begin{equation}\label{eq:SRB-T(AG)}
\langle T^{(A,G)}(L)\rangle=\sum_{n=1}^{2}
\langle T_M(L/L^{(A,G)(b)}_n)\rangle,
\end{equation}
that ignores this contribution. In Fig.~\ref{fig:SRBarrayVSkdOverPi} (f) the first 
and second propagating modes are in ballistic transport regime. When increasing 
$kd/\pi$, this is portrayed by the sharp increase of the transmittance and, then, by 
the ballistic behavior given by the asymptote \eqref{eq:SRB-TM-bal}. This 
step-like behavior indicates the openness of the channels with almost unit 
transmittance and, consequently, the total transmittance given by 
Eq.~\eqref{eq:SRB-T-Bal}, as the total number of propagating modes. Since in 
frame (f) the main contribution is due to the amplitude-scattering mechanism, the 
behavior of the total transmittance follows the approximation 
\eqref{eq:SRB-T(AG)} (see the overlapping of curves related to 
Eqs.~\eqref{eq:SRB-Twn} and \eqref{eq:SRB-T(AG)}).

%%%%%%%%%%%%%%%%%%%%%%%%%%%%%%%%%%%%%%%%%%%%%%%%%%%%%%%%%%%%%%%%%%%%%%%%%%%%%%%
\section{Two-mode ASRB Waveguide}
\label{sec:ASRB}
%%%%%%%%%%%%%%%%%%%%%%%%%%%%%%%%%%%%%%%%%%%%%%%%%%%%%%%%%%%%%%%%%%%%%%%%%%%%%%%

In contrast with the SRB waveguide, the theory for the ASRB waveguide can not 
describe the whole transition from ballistic to localized transport, which is 
expected for large enough $L$ (numerical evidence of this transition can be 
found, for instance, in 
Refs.~\onlinecite{GrcTrrSnzNto_1997,SncFrlYrkMrd_1998}).
However, the analysis of the mode-attenuation lengths that is performed here for 
two mode waveguides, predicts many of important characteristics of transport. 
Thus, as is remarked in Sec.~\ref{sec:Intro}, the attenuation in the ASRB 
waveguide is determined by the interplay between the gradient-scattering and 
square-gradient-scattering mechanisms, i.e., it is determined by the interplay 
between inter-mode and intra-mode scattering. Below, firstly, a more detailed 
analysis of mode-attenuation lengths allow us to compute the interval of 
parameters in which only one of these mechanisms prevails, as well as their 
behavior inside that interval. Secondly, the theory already developed for the SRB 
waveguide, in which the transport properties of the $n$th channel are specified 
by the scaling parameter $L/L_n^{(b)}$, is applied when only the intra-mode 
scattering governs the attenuation and the channels can be considered as 
independent. Thirdly, the theory that accounts for the transport properties of the 
ASRB waveguide, when the inter-mode scattering determines the attenuation 
(when the channels become mixed), is developed. Finally, a small region of 
transition, where no analytical results exist, is only indicated.

%%%%%%%%%%%%%%%%%%%%%%%%%%%%%%%%%%%%%%%%%%%%%%%%%%%%%%%%%%%%%%%%%%%%%%%%%%%%%%%
\subsection{ASRB: Competition between surface-scattering mechanisms}
%%%%%%%%%%%%%%%%%%%%%%%%%%%%%%%%%%%%%%%%%%%%%%%%%%%%%%%%%%%%%%%%%%%%%%%%%%%%%%%

The competition between surface-scattering mechanisms is again discussed 
within the two intervals of correlation length $\Rc$, which are denoted as the 
region of \emph{small-scale roughness} ($k\Rc\ll1$) and the region of 
\emph{large-scale roughness} ($k\Rc\gg1$). Both regions correspond to weak 
correlations between successive reflections of the wave from rough boundaries 
($\Rc\ll\Lambda_n$). See details about the definition of the regions of correlation 
length in subsection~\ref{subsec:SRBCompetition}.

%%%%%%% R1.
\emph{\bf Small-scale-roughness region - } In this region, see 
Eq.~\eqref{eq:region1_SSR}, if the correlation length $\Rc$ is smaller than the 
\emph{crossing point} $\Rc_{\cpt n}$, the first term in 
Eqs.~\eqref{eq:ASRB_L1(b)(f)_L2(b)(f)} can be neglected. Then, after the 
substitution of Eq.~\eqref{eq:GaussianT_SSR} into the second term of 
Eqs.~\eqref{eq:ASRB_L1(b)(f)_L2(b)(f)} one arrives to 
$1/L_n^{(b)}\approx 1/L_n^{(f)}$, and the following asymptotic formulas for the 
backscattering lengths are obtained
\begin{subequations}\label{eq:ASRB_SSRApprox1_Ln(b)}
\begin{eqnarray}
\label{eq:ASRB_SSRApprox1_L1(b)}
\frac{1}{L_{1}^{(b)}}&\approx&64.7\;k_1\,
\frac{(\sigma/d)^4}{(k_1 R)^3} \nonumber \\ 
&& \qquad\qquad \text{for}\;k\Rc\ll k\Rc_{\cpt1}\ll1,\\
\label{eq:ASRB_SSRApprox1_L2(b)}
\frac{1}{L_{2}^{(b)}}&\approx& 1035.9\;k_2\,
\frac{(\sigma/d)^4}{(k_2 R)^3} \nonumber \\
&& \qquad\qquad \text{for}\; k\Rc\ll k\Rc_{\cpt2}\ll 1. 
\end{eqnarray}
\end{subequations}
It is necessary to be aware that the present analysis of the competition is valid for 
the values of mode parameter $kd/\pi$ not very close to the resonances 
$kd/\pi=1,2$. Note particularly, that the approximations 
\eqref{eq:ASRB_SSRApprox1_Ln(b)} do not work at $kd/\pi=2$, since this 
resonance is included in the neglected terms $1/L^{(A,G)(b)}_n$ and 
$1/L^{(A,G)(f)}_n$.

If $\Rc$ is larger than $\Rc_{\cpt n}$, the square-gradient term in 
Eqs.~\eqref{eq:ASRB_L1(b)(f)_L2(b)(f)} can be neglected. Then, after the 
substitution of \eqref{eq:GaussianW_SSR} into the first term, the 
mode-attenuation lengths \eqref{eq:Ln(f)(b)} are approximated to
\begin{subequations}\label{eq:SSRApprox2_Ln}
\begin{eqnarray}
\label{eq:SSRApprox2_L1}
\frac{1}{L_{1}}&\approx& 80.2\; k_2\,
\frac{(\sigma/d)^2 (k_1\Rc)}{(k_1 d/\pi)^2 (k_2 d/\pi)^2} \nonumber \\
&& \qquad\qquad \text{for}\; k\Rc_{\cpt1}\ll k\Rc\ll 1, \\
\label{eq:SSRApprox2_L2}
\frac{1}{L_{2}}&\approx& \frac{1}{L_{1}} \qquad \text{for}\; k\Rc_{\cpt2}\ll k\Rc\ll 1.
\end{eqnarray}
\end{subequations}

The dimensionless crossing point $k\Rc_{\cpt n}$ of the $n$th length can be 
located either on the border between the regions of small- and 
large-scale-roughness, or inside the first region. In the former case, 
$k\Rc_{\cpt n}\sim 1$, and within the whole small-scale roughness region the 
mode-attenuation lengths are described only by square-gradient-scattering 
contribution, i.e., by Eq.~\eqref{eq:ASRB_SSRApprox1_Ln(b)}. In the latter case, 
$k\Rc_{\cpt1}$ and $k\Rc_{\cpt2}$ are computed by searching for the intersection 
of the asymptotes \eqref{eq:ASRB_SSRApprox1_Ln(b)} with 
\eqref{eq:SSRApprox2_Ln}. Thus, the crossing points read
\begin{subequations}
\begin{eqnarray}\label{eq:ASRB_kR_c_n}
k\Rc_{\cpt 1}
&\approx& 0.36\; \sqrt[4]{\frac{k_2}{k_1}}\; \sqrt{\frac{\sigma}{d}}\; k d, \\
k\Rc_{\cpt 2}
&\approx& 0.72\; \sqrt[4]{\frac{k_1}{k_2}}\; \sqrt{\frac{\sigma}{d}}\; k d.
\end{eqnarray}
\end{subequations}

%%%%%%% R2.
\emph{\bf Large-scale-roughness region - } In this region, see
Eq.~\eqref{eq:region2_LSR_WC}, the square-gradient terms in 
Eqs.~\eqref{eq:ASRB_L1(b)(f)_L2(b)(f)} can be neglected and the 
mode-attenuation lengths \eqref{eq:Ln(f)(b)} are approximated to
\begin{subequations}\label{eq:LSRWCApprox_Ln}
\begin{eqnarray}
\label{eq:LSRWCApprox_L1}
\frac{1}{L_{1}}&\approx& 40.1\;k_2\,
\frac{(\sigma/d)^2 (k_1\Rc)} {(k_1d/\pi)^2(k_2d/\pi)^2}\nonumber\\
&\times&\left\{\exp\left[-\frac{(k_1+k_2)^2 R^2}{2}\right]\right. \nonumber\\
&&+\left. \exp\left[-\frac{(k_1-k_2)^2 R^2}{2}\right]\right\},\nonumber \\
&&\qquad\qquad \text{for}\;1\ll k\Rc\ll k\Lambda_1,\\
\label{eq:LSRWCApprox_L2}
&&\frac{1}{L_{2}}\approx \frac{1}{L_{1}},
\qquad\text{for}\;1\ll k\Rc\ll k\Lambda_2.
\end{eqnarray}
\end{subequations}

%%%%%%%%%%%%%%%%%%%%%%%%%%%%%%%%%%%%%%%%%%%%%%%%%%%%%%%%%%%%%%%%%%%%%%%%%%%%%%%
\subsection{ASRB: Total and mode transmittances}
%%%%%%%%%%%%%%%%%%%%%%%%%%%%%%%%%%%%%%%%%%%%%%%%%%%%%%%%%%%%%%%%%%%%%%%%%%%%%%%

The preceding analysis of mode-attenuation lengths states that values of 
correlation length smaller than the first crossing point, $\Rc<\Rc_{c1}$, result in 
\emph{both} mode-attenuation lengths governed by intra-mode scattering only, 
see Eqs.~\eqref{eq:ASRB_SSRApprox1_Ln(b)}. Therefore, mode and total 
transmittances are given by Eqs.~\eqref{eq:SRB-TM-def} and 
\eqref{eq:SRB-Twn}, however, with mode-backscattering lengths $L_1^{(b)}$ 
and $L_2^{(b)}$ given by Eqs.~\eqref{eq:ASRB_L1(b)} and 
\eqref{eq:ASRB_L2(b)} that are well described by their asymptotes 
\eqref{eq:ASRB_SSRApprox1_L1(b)} and \eqref{eq:ASRB_SSRApprox1_L2(b)}.

Also the preceding analysis says that values of correlation length
larger than the second crossing point, $\Rc_{c2}<\Rc$, result in
mode-attenuation lengths associated with inter-mode scattering only, see 
Eqs.~\eqref{eq:SSRApprox2_Ln}) and \eqref{eq:LSRWCApprox_Ln}. In this case 
the theory that accounts for the transport properties should differ from that used in 
SRB waveguide. Within the approach presented here, the average two-particle 
Green's function entering in Eq.~\eqref{eq:T(L)-def}, for the transmittance, is 
assumed to be equal to the product of two average one-particle ones. Therefore, 
to obtain the total transmittance $\langle T(L)\rangle$, we can substitute 
Eq.~\eqref{eq:avGF} into Eq.~\eqref{eq:T(L)-def}. After evaluation of the integrals, 
the following result becomes apparent
\begin{equation}\label{eq:ASRB-Twn}
\langle T(L)\rangle=\sum_{n=1}^{N_d} \langle T_M(L/L_n)\rangle,
\quad \text{for}\quad \Rc_{c2}<\Rc.
\end{equation}
Here the total transmittance $\langle T(L)\rangle$ reads as a sum of average 
mode transmittances $\langle T_M(L/L_n)\rangle $, describing the transparency 
of every $n$th propagating mode. Within the assumption necessary for the 
averaging procedure to be reasonable, they are determined by the following 
expression
\begin{equation}\label{eq:ASRB-TM-def}
\langle T_M(L/L_n)\rangle=2\,\frac{L_n}{L}\left\{1-\frac{L_n}{L}
\left[1-\exp\left(-\frac{L}{L_n}\right)\right]\right\}.
\end{equation}
One can see that the mode transmittance is described by a function that depends 
on the parameter $L/L_n$ only. Note that in accordance with 
Eq.~\eqref{eq:Ln(f)(b)}, the above parameter depends on {\it both} backward and 
forward scattering lengths explicitly given in 
Eqs.~\eqref{eq:ASRB_L1(b)(f)_L2(b)(f)} and obeying the asymptotic expressions 
\eqref{eq:LSRWCApprox_Ln}.

\begin{figure}
\includegraphics[width=8cm]{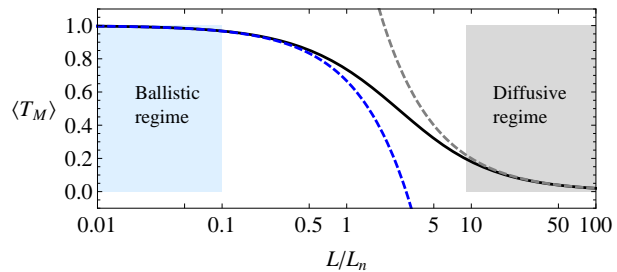}
\caption{\label{fig:ASRBTMandAsymptotes}
(Color online) ASRB waveguide: Mode transmittance described by 
Eq.~\eqref{eq:ASRB-TM-def} (solid line) and its asymptotes, given by 
\eqref{eq:ASRB-TM-bal} and \eqref{eq:ASRB-TM-dif} for ballistic and diffusive 
regimes (dashed lines). The region where the diffusion terminates due to the 
influence of localization, is shown schematically.}
\end{figure}

The mode transmittance \eqref{eq:ASRB-TM-def} exhibits the ballistic behavior 
for large mode attenuation length,
\begin{equation}\label{eq:ASRB-TM-bal}
\langle T_M(L/L_n)\rangle\approx1-L/3L_n\quad\text{for}\quad L\ll L_n.
\end{equation}
In this case the $n$th conducting channel is practically transparent. On the 
contrary, the mode transmittance is small when $L_n$ turns out to be much less 
than the length of the waveguide,
\begin{equation}\label{eq:ASRB-TM-dif}
\langle T_M(L/L_n)\rangle\approx2L_n/L\quad\text{for}\quad L_n\ll L.
\end{equation}
The inverse proportional dependence of Eq.~\eqref{eq:ASRB-TM-dif} on the 
waveguide length $L$ implies the diffusive regime of wave transport at a given 
$n$th mode. Figure~\ref{fig:ASRBTMandAsymptotes} shows the behavior of
Eq.~\eqref{eq:ASRB-TM-def} and its asymptotes \eqref{eq:ASRB-TM-bal} and 
\eqref{eq:ASRB-TM-dif} versus the parameter $L/L_n$.

It is necessary now to consolidate both aforementioned complementary 
approaches. Firstly, when $\Rc<\Rc_{\cpt1}$, the intra-mode scattering gives 
rise to the hierarchy of mode-backscattering lengths, $L_2^{(b)}<L_1^{(b)}$, see 
Eqs.~\eqref{eq:ASRB_SSRApprox1_L1(b)} and 
\eqref{eq:ASRB_SSRApprox1_L2(b)}. Due to this hierarchy and in accordance 
with two mode-transport regimes \eqref{eq:SRB-TM-bal} and 
\eqref{eq:SRB-TM-loc}, the total transmittance \eqref{eq:SRB-Twn} can exhibit 
the ballistic, coexistence and localized transport regimes. Secondly, when 
$\Rc_{\cpt2}<\Rc$, in accordance with two mode-transport regimes 
\eqref{eq:ASRB-TM-bal} and \eqref{eq:ASRB-TM-dif}, the total transmittance 
\eqref{eq:ASRB-Twn} can exhibit the ballistic and diffusive regimes. Here the 
coexistence transport regime can not emerge because the inter-mode scattering 
results in mode-attenuation lengths that approximate each other, 
$L_1\approx L_2$, see \eqref{eq:SSRApprox2_L2} and 
\eqref{eq:LSRWCApprox_L2}. Specifically, the transport regimes of the 
\emph{total} transmittance are defined as follows:

\begin{enumerate}
\item[(i)]
In the \emph{regime of localization}, when the largest mode-backscattering 
length $L_1^{(b)}$ is smaller than the waveguide length $L$, and the correlation 
length $\Rc$ is smaller than the first crossing point $\Rc_{\cpt1}$,
\begin{equation}\label{eq:ASRB_LocRegime}
1<L/L_1^{(b)}<L/L_2^{(b)},\quad \Rc<\Rc_{\cpt1},
\end{equation}
both propagating modes are strongly localized and their transmittances are 
exponentially small. The total transmittance is approximately equal to the 
transmittance of the first mode obeying the asymptote \eqref{eq:SRB-TM-loc},
\begin{equation}\label{eq:ASRB-T-loc}
\begin{split}
\langle T(L)\rangle&\approx\langle T_M(L/L_1^{(b)})\rangle\\
&\approx\frac{\pi^{5/2}}{16}\left(\frac{L}{4L_1^{(b)}}\right)^{-3/2}
\exp\left(-\frac{L}{4L_1^{(b)}}\right),
\end{split}
\end{equation}
with $L_1^{(b)}$ given by Eq.~\eqref{eq:ASRB_SSRApprox1_L1(b)}. The 
waveguide is non-transparent in this regime.
\item[(ii)]
The \emph{coexistence regime} arises when the smallest backscattering length 
$L_2^{(b)}$ is smaller, while the largest backscattering length $L_1^{(b)}$ is 
larger than the waveguide length $L$, with the correlation length $\Rc$ smaller 
than the first crossing point $\Rc_{c1}$,
\begin{equation}\label{eq:ASRB-CoexRegime}
L/L_1^{(b)}<1<L/L_2^{(b)},\quad \Rc<\Rc_{c1}.
\end{equation}
In this case, the first mode manifests the ballistic behavior \eqref{eq:SRB-TM-bal}, 
while the second mode is still localized in line with Eq.~\eqref{eq:SRB-TM-loc}. 
Therefore, as before, the total transmittance \eqref{eq:SRB-Twn} is determined 
by the transmittance of the first mode, however governed by the ballistic 
asymptote \eqref{eq:SRB-TM-bal},
\begin{equation}\label{eq:ASRB-T-Int}
\langle T(L)\rangle\approx\langle T_M(L/L_1^{(b)})\rangle\approx1.
\end{equation}
The coexistence of the ballistic and diffusive regimes is possible when the 
hierarchy of the mode-attenuation lengths, $L_2^{(b)}<L_1^{(b)}$, occurs.
\item[(iii)]
The \emph{diffusive regime} emerges when both mode transmittances have the 
diffusive behavior and when the correlation length $R$ is larger than the second 
crossing point $\Rc_{c2}$,
\begin{equation}\label{eq:ASRB-DifRegime}
1<L/L_1\approx L/L_2, \quad \Rc_{\cpt2}<\Rc.
\end{equation}
This regime arises if the attenuation is caused by the gradient-scattering 
mechanism. The mode transmittances have approximately the same behavior 
that is described by Eq.~\eqref{eq:ASRB-TM-dif} and, as a consequence, the 
total transmittance can be expressed as a double value of one of them,
\begin{subequations}\label{eq:ASRB-GS-T1T2T-dif}
\begin{eqnarray}
\label{eq:ASRB-GS-T1T2-dif}
\langle T_M(L/L_1)\rangle &\approx& \langle T_M(L/L_2)\rangle, \\
\label{eq:ASRB-GS-T-dif}
\langle T(L)\rangle &\approx& 2\langle T_M(L/L_1)\rangle\approx 4L_1/L.
\end{eqnarray}
\end{subequations}
Here the approximate expression for $L_1$ is given either by
Eq.~\eqref{eq:SSRApprox2_Ln} if $k\Rc_{\cpt 2}\ll k\Rc<1$, or by 
Eq.~\eqref{eq:LSRWCApprox_Ln} if $1\ll k\Rc\ll k\Lambda_n$.
\item[(iv)]
The \emph{ballistic regime} emerges either under the conditions, %
\begin{subequations}\label{eq:ASRB-BalRegime}
\begin{eqnarray}
\label{eq:ASRB-BalRegime1}
&&L/L_1^{(b)}<L/L_2^{(b)}<1\quad\text{for}\;\Rc<\Rc_{\cpt1}, \;\text{or}\\ 
\label{eq:ASRB-BalRegime2}
&&L/L_1\approx L/L_2<1\quad\text{for}\quad\Rc_{\cpt2}<\Rc,
\end{eqnarray}
\end{subequations}
when the smallest mode-attenuation length, $L_2^{(b)}$ or $L_2$, is larger than 
$L$. Both conducting channels are open having approximately unit 
transmittances, \eqref{eq:SRB-TM-bal} or \eqref{eq:ASRB-TM-bal}. The total 
transmittance approaches the total number of propagating modes,
\begin{equation}\label{eq:ASRB-T-Bal}
\langle T(L)\rangle\approx2.
\end{equation}
Here the waveguide has almost perfect transparency.
\end{enumerate}

Note that our model is no able to describe the transmittances for a 
\emph{transition} region that emerges within the interval
$\Rc_{\cpt1}<\Rc<\Rc_{\cpt2}$, when both intra-mode and inter-mode scattering 
have a comparable weight. However, this region is actually small in comparison 
with the large range of values of $\Rc$ for which the analytical approach has 
been developed.

%%%%%%%%%%%%%%%%%%%%%%%%%%%%%%%%%%%%%%%%%%%%%%%%%%%%%%%%%%%%%%%%%%%%%%%%%%%%%%%
\subsection{ASRB: Transmittance vs correlation length}
%%%%%%%%%%%%%%%%%%%%%%%%%%%%%%%%%%%%%%%%%%%%%%%%%%%%%%%%%%%%%%%%%%%%%%%%%%%%%%%

\begin{figure}
\includegraphics[width=\columnwidth]{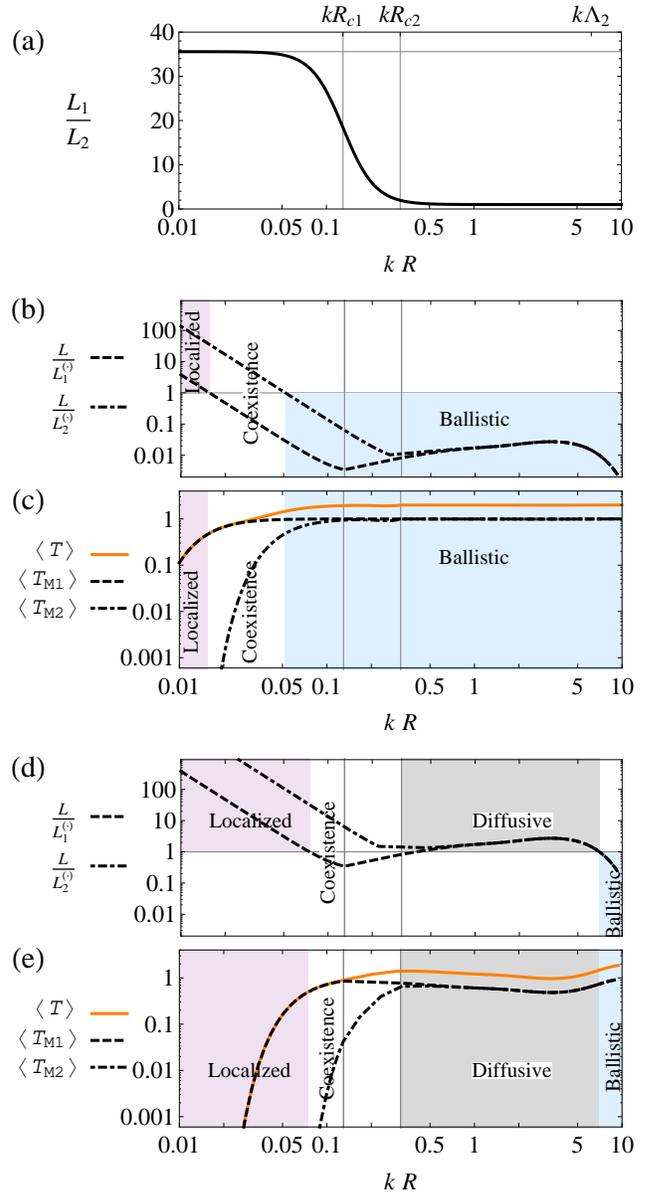}
\caption{\label{fig:ASRBarrayVSkR}
(Color online) ASRB waveguide: Mode-backscattering lengths, mode-attenuation 
lengths, $L_n^{(\cdot)}$, and transmittances, 
$\langle T_{Mn}\rangle=\langle T_M(L/L_n^{(\cdot)})\rangle$,
$\langle T(L)\rangle$, vs dimensionless correlation length $k\Rc$
(here $L_n^{(\cdot)}=L_n^{(b)}$ for $k\Rc<k\Rc_{\cpt1}$ and
$L_n^{(\cdot)}=L_n$ for $k\Rc_{\cpt2}<k\Rc$). The crossing points 
$k\Rc_{\cpt1}\approx 0.13$ and $k\Rc_{\cpt2}\approx 0.32$ are indicated on the 
top of the upper plot. Note that there is no analytical results  within the transition 
region $k\Rc_{\cpt1}<k\Rc<k\Rc_{\cpt2}$, in which the curves shown are 
obtained by simple interpolation. The global parameters are $\sigma/d=0.0025$ 
and $kd/\pi=2.54$, ($k\sigma=0.02$).
In (a) the ratio $L_1/L_2$ is depicted; the asymptotes, $L_1/L_2\approx35.78$ 
and $L_1/L_2\approx1$ can be noted.
In (b)-(c) the normalized waveguide length is $L/d=160$; the boundary between 
the localized and coexistence regimes is at $k\Rc\approx 0.016$, then, the 
boundary between the coexistence and ballistic regimes is at 
$k\Rc\approx 0.05$. In (d)-(e) $L/d=16,000$: the boundary between the localized 
and coexistence regimes is at $k\Rc\approx 0.075$, then, the boundary between 
the diffusive and ballistic regimes is at $k\Rc\approx 7.0$.}
\end{figure}

In order to discuss the dependence of the transmittances and their scaling 
parameters upon the dimensionless correlation length $k\Rc$, 
Fig.~\ref{fig:ASRBarrayVSkR} is presented. In the region $k\Rc<k\Rc_{\cpt1}$, 
where intra-mode scattering prevails, the transmittances 
$\langle T_M(L/L_1^{(b)})\rangle$, $\langle T_M(L/L_2^{(b)})\rangle$, 
$\langle T(L)\rangle$, the scaling parameters $L/L_1^{(b)}$,  $L/L_2^{(b)}$, and 
the ratio of these parameters $L_1^{(b)}/L_2^{(b)}$, are depicted. Otherwise, in 
the region $k\Rc_{\cpt2}<k\Rc$, where inter-mode scattering prevails, the 
transmittances $\langle T_M(L/L_1)\rangle$, $\langle T_M(L/L_2)\rangle$, 
$\langle T(L)\rangle$, the scaling parameters $L/L_1$,  $L/L_2$, and the ratio 
$L_1/L_2$, are depicted. In the first region the transmittances are computed from 
Eqs.~\eqref{eq:SRB-Twn} and \eqref{eq:SRB-TM-def}, whereas in the second 
region, from Eqs.~\eqref{eq:ASRB-Twn}, \eqref{eq:ASRB-TM-def}. In both 
ranges, however, the scattering lengths are computed from 
Eqs.~\eqref{eq:ASRB_L1(b)(f)_L2(b)(f)}.

If $\Rc$ decreases, the lengths $L_n$ and $L_n^{(b)}$ decreases as $\Rc^3$ 
whereas the cycle length $\Lambda_n$ has a fixed value, thus, the smallest 
value of $\Rc$ is restricted by the criterion \eqref{eq:WeakSurfaceScattCriteria1}; 
here this criterion is fulfilled since $\Lambda_1/2L_1^{(b)}\approx 0.03 \ll 1$ and 
$\Lambda_2/2L_2^{(b)}\approx 0.33 \ll 1$ at the left boundary of the plots, 
$k\Rc=0.01$. Because the plots illustrate the case of weak correlations 
($\Rc\ll \Lambda_n$), the condition
\eqref{eq:WeakSurfaceScattCriteria2} is satisfied automatically if 
\eqref{eq:WeakSurfaceScattCriteria1} is met.

The ratio of the lengths $L_1^{(b)}/L_2^{(b)}$, which is equal to the ratio 
$L_1/L_2$ for $k\Rc<k\Rc_{\cpt1}$, is in Fig.~\ref{fig:ASRBarrayVSkR} (a). It 
exhibits the crossover from square-gradient scattering to gradient scattering. 
Therefore, with the increase of $k\Rc$, the ratio decreases from the large value
\begin{equation}\label{eq:AprxR1_L1_over_L2_ASym}
\frac{L_1^{(b)}}{L_2^{(b)}}\approx\frac{L_1}{L_2}\approx16 
\left(\frac{k_1}{k_2}\right)^2
\quad\text{for}\quad k\Rc\ll k\Rc_{\cpt 1},
\end{equation}
to the unity,
\begin{equation}\label{eq:AprxR2_L1_over_L2_ASym}
\frac{L_1}{L_2}\approx1\quad\text{for}\quad
k\Rc_{\cpt 2}\ll k\Rc\ll k\Lambda_2.
\end{equation}
The asymptotes \eqref{eq:AprxR1_L1_over_L2_ASym} and
\eqref{eq:AprxR2_L1_over_L2_ASym} are obtained, respectively, through 
Eqs.~\eqref{eq:ASRB_SSRApprox1_Ln(b)} and \eqref{eq:LSRWCApprox_Ln}. 
In contrast with the SRB waveguide, here the equality of mode-attenuation 
lengths, $L_1\approx L_2$, emerges as a result of a strong intermode mixing, 
which is governed by the gradient-scattering mechanism. This non-trivial fact is 
in a strong contrast with a non-isotropic character of surface scattering in the 
channel space \cite{SncFrlYrkMrd_1998,GrcSnzNto_PhysRevLett_2000}.

In Fig.~\ref{fig:ASRBarrayVSkR} (b)-(c) and (d)-(e) two cases are illustrated in 
which, depending upon the roughness correlation length $k\Rc$, different 
transport regimes can arise. The first case corresponds to a moderately large 
waveguide that can present the localized, coexistence or ballistic transport 
regime; see frames (b) and (c). In comparison, the second case corresponds to a 
larger waveguide that can present the diffusive transport regime in addition to the 
aforementioned ones; see frames (d) and (e). Note that different transport 
regimes, defined in Eqs.~\eqref{eq:ASRB_LocRegime}, 
\eqref{eq:ASRB-CoexRegime}, \eqref{eq:ASRB-DifRegime} and 
\eqref{eq:ASRB-BalRegime}, can be graphically identified with the use of frames 
(b) and (d) for the scaling parameters.

The total and mode transmittances are plotted in Figs.~\ref{fig:ASRBarrayVSkR} 
(c) and (e). These plots demonstrate that, due to the hierarchy of 
mode-backscattering lengths, $L_2^{(b)}<L_1^{(b)}$, the transmittance of the 
second mode is always smaller than the transmittance of the first one,
\begin{equation}\label{eq:ASRB-T2<T1}
\langle T_M(L/L_2^{(b)})\rangle<\langle T_M(L/L_1^{(b)})\rangle.
\end{equation}
Therefore, within the localization \eqref{eq:ASRB_LocRegime} and coexistence 
\eqref{eq:ASRB-CoexRegime} regions, the total transmittance 
\eqref{eq:SRB-Twn} is mainly contributed by the first term, i.e. by the first-mode 
transmittance,
\begin{equation}\label{eq:ASRB-T=T1}
\langle T(L)\rangle\approx\langle T_M(L/L_1^{(b)})\rangle.
\end{equation}
Within the localization regime, $\langle T(L)\rangle$ behaves as indicated in 
Eq.~\eqref{eq:ASRB-T-loc}. Within the coexistence regime the total transmittance 
is described by Eq.~\eqref{eq:ASRB-T-Int}. In the diffusive regime the mode 
transmittances are equal, and behave as indicated in 
Eq.~\eqref{eq:ASRB-TM-dif}. The total transmittance has twice their value, see 
Fig.~\ref{fig:ASRBarrayVSkR} (e) and Eqs.~\eqref{eq:ASRB-GS-T1T2T-dif}. For 
the ballistic regime \eqref{eq:ASRB-BalRegime2}, both conducting channels are 
open
\begin{equation}\label{eq:ASRB-T1T2-Bal}
\langle T_M(L/L_1^{(b)})\rangle\approx 
\langle T_M(L/L_2^{(b)})\rangle\approx1.
\end{equation}
Here the approximation given by Eq.~\eqref{eq:ASRB-TM-bal} describes well the 
behavior of the mode transmittances. Correspondingly, the value of the total 
transmittance tends to the total number $N_d=2$ of the propagating modes, see 
Eq.~\eqref{eq:ASRB-T-Bal}.

Note that the coexistence transport regime arises if the length $L$ is such that 
$L_2^{(b)}<L<L_1^{(b)}$, meanwhile the implied hierarchy $L_2^{(b)}<L_1^{(b)}$ 
emerges if the correlation length is smaller than the crossing point, 
$k\Rc\ll k\Rc_{\cpt 1}$. Therefore, in the ASRB waveguide the scattering 
governed by the square-gradient-scattering mechanism is a necessary condition 
for the coexistence transport regime to occur. In comparison, if the 
gradient-scattering mechanism prevails, the mode-attenuation lengths are almost 
equal, $L_1\approx L_2$. Then, with a continuous increment of $L$, the system 
passes directly from the ballistic to the diffusive regime.

%%%%%%%%%%%%%%%%%%%%%%%%%%%%%%%%%%%%%%%%%%%%%%%%%%%%%%%%%%%%%%%%%%%%%%%%%%%%%%%
\subsection{ASRB: Transmittance vs mode parameter}
%%%%%%%%%%%%%%%%%%%%%%%%%%%%%%%%%%%%%%%%%%%%%%%%%%%%%%%%%%%%%%%%%%%%%%%%%%%%%%%

\begin{figure}
\includegraphics[width=\columnwidth]{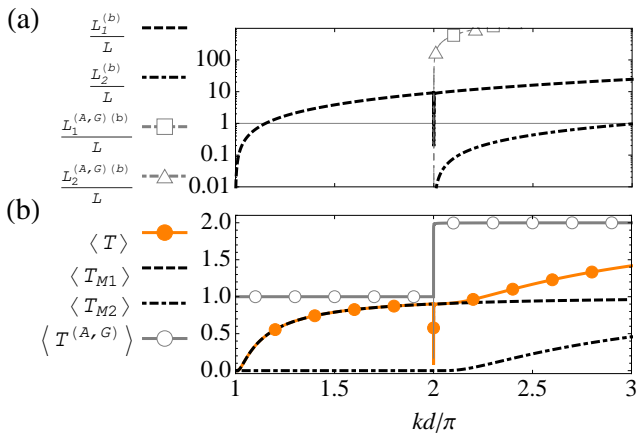}
\caption{\label{fig:ASRBarrayVSkdOverPi}
(Color online) ASRB waveguide: Mode-backscattering lengths and 
transmittances versus the mode parameter for $\sigma/d=0.0025$,
$\Rc/d\approx0.005$ and $L/d=160$. Note that for fixed value of the correlation 
length $\Rc$, and mode parameter within the interval $2<kd/\pi<3$, the 
dimensionless correlation length $k\Rc$ ranges 
$0.031\lessapprox k\Rc\lessapprox0.047$, therefore, the data correspond to the 
small-scale roughness \eqref{eq:region1_SSR} and the coexistence regime 
\eqref{eq:ASRB-CoexRegime}.}
\end{figure}

The mode-attenuation lengths and transmittances  as functions of the mode 
parameter $kd/\pi$ are presented in Fig.~\ref{fig:ASRBarrayVSkdOverPi} within 
the range of first two modes. The value of the parameters are selected in order to 
illustrate, within the range $2<kd/\pi<3$, the coexistence transport regime of the 
ASRB waveguide, which is defined by Eq.~\eqref{eq:ASRB-CoexRegime}.

Fig.~\ref{fig:ASRBarrayVSkdOverPi} (a) shows the plots of $L_n^{(b)}/L$ and 
$L_n^{(A,G)(b)}/L$. Here $L_n^{(b)}$ is computed from 
Eqs.~\eqref{eq:ASRB_L1(b)} and \eqref{eq:ASRB_L2(b)}, whereas
$L_n^{(A,G)(b)}$ is given by the first term of the above equations, i.e.,
\begin{equation}\label{eq:Ln(SG)}
\begin{split}
&\frac{1}{L_1^{(A,G)(b)}}=\frac{1}{L_2^{(A,G)(b)}}\\
&=40.1\; \frac{(\sigma/d)^2 (k_1\Rc)\,k_2}{(k_1 d/\pi)^2 (k_2 d/\pi)^2}
\exp\left[\frac{-(k_1+k_2)^2 R^2}{2}\right].\nonumber\\
\end{split}
\end{equation}

If $kd/\pi$ increases, all the attenuation lengths also increase but exhibit 
characteristic dips at $kd/\pi=1,2$ where a new normal mode opens. As we 
already noted, we can only indicate the position of these dips and do not describe 
their line shape in the narrow vicinity of integer values $kd/\pi$ (see short 
discussion of these resonances in connection with 
Fig.~\ref{fig:SRBarrayVSkdOverPi}). In contrast with the SRB waveguide, the 
curves depicting the first propagating mode of the ASRB waveguide show a dip 
at $kd/\pi=2$. This dip is caused by the inter-mode scattering from the first 
propagating mode into the second, and this scattering is governed by the 
gradient-scattering mechanism. Remarkably, far from the resonances at 
$kd/\pi=1,2$, the lengths $L^{(A,G)(b)}_1/L$ and $L^{(A,G)(b)}_2/L$ are larger 
than the mode-backscattering lengths, $L_1^{(b)}/L$ and $L_2^{(b)}/L$. This is 
because of the main contribution of the square-gradient-scattering lengths, 
$L^{(SG)(b)}_1/L$ and $L^{(SG)(b)}_2/L$. Therefore, the approximate behavior 
of $L_n^{(b)}/L$ is given by Eqs.~\eqref{eq:ASRB_SSRApprox1_L1(b)} and 
\eqref{eq:ASRB_SSRApprox1_L2(b)}(not plotted in the Figure).

The total transmittance \eqref{eq:SRB-Twn} and mode transmittances
\eqref{eq:SRB-TM-def} are plotted in Fig.~\ref{fig:ASRBarrayVSkdOverPi} (b). 
Note that the dips pertaining to the curves of $L_1$ and $L_2$ become evident 
in the curves for mode transmittances $\langle T_M(L/L_1^{(b)})\rangle$ and 
$\langle T_M(L/L_2^{(b)})\rangle$. Certainly, the curve for the total transmittance 
$\langle T(L)\rangle$ incorporates all the dips of the mode transmittances. 
Remarkably, without the weight of the square-gradient scattering mechanism, 
both mode transmittances would have the same value and the total transmittance 
would have ballistic behavior. This fact can be graphically observed with the help 
of the curve related to the total transmittance given in Eq.~\eqref{eq:SRB-T(AG)}, 
in which the square-gradient-scattering mechanism has been ignored.

%%%%%%%%%%%%%%%%%%%%%%%%%%%%%%%%%%%%%%%%%%%%%%%%%%%%%%%%%%%%%%%%%%%%%%%%%%%%%%%
\section{Conclusions}
\label{sec:Conclusions}
%%%%%%%%%%%%%%%%%%%%%%%%%%%%%%%%%%%%%%%%%%%%%%%%%%%%%%%%%%%%%%%%%%%%%%%%%%%%%%%

In this paper we report new results for transport properties of quasi-1D 
surface-disordered waveguides (or electronic conductors). Two models with 
symmetric and antisymmetric rough boundaries were studied in great detail. By 
examining these models, one passes from a system in which the attenuation of 
the propagating modes is associated with solely intra-mode scattering (the SRB
waveguide), to a system in which both intra- and inter-mode scattering determine 
its attenuation (the ASRB waveguide). The comparison between the two models 
allows us to elucidate the role of all surface-scattering mechanisms.

As was shown in our previous analytical studies \cite{IzrMkrRnd_PRB_2005,
IzrMkrRnd-PRB_2006,RndIzrMkr-MSMW-2007,RndIzrMkr-PRB-2007,
RndIzrMkr-MMET-2008}, one of mechanisms of scattering (the so-called 
square-gradient mechanism) is typically neglected in the literature since it 
emerges in the second order of perturbation theory. 
Specifically, the corresponding term in the expression for the inverse 
attenuation length is proportional to the squared variance of disorder, $\sigma^4$, 
and for this reason the square-gradient terms seem to be much smaller than the 
others describing conventional mechanisms of scattering (giving rise to the 
$\sigma^2$-dependence). However, apart from the dependence on the disorder 
strength, one has to take into account the correlation properties of scattering 
profiles. It was found that the square-gradient mechanism essentially depends on 
the correlations of different nature as compared with the $\sigma^2$-terms in the 
total expression for the inverse attenuation length.

In particular, it was argued that in some regions of parameters, the 
square-gradient terms can prevail the $\sigma^2$-terms and, therefore, have to 
be considered as well, in spite of their apparent smallness. These predictions 
have been carefully analyzed in this paper, within two models with symmetric and 
antisymmetric scattering profiles. Due to a principal difference between the two 
models, we analyze them separately.

For the SRB waveguide the transmission occurs along two modes 
independently, due to the underlying symmetry of the model. Therefore, the 
transport through each of two modes can be described in terms of standard 
theories of localization in one-dimensional models with random potentials. Thus, 
the total transmission is the sum of partial transmissions with the corresponding 
quantized transverse momentum. Our results demonstrate that in this case a 
typical situation occurs for which the transmission along two channels can be 
very different due to different values of corresponding backscattering lengths.

For the ASRB waveguide there exist a competition of two surface-scattering 
mechanism, which in contrast with the SRB waveguide, induce the interplay 
between intra-mod and inter-mode scattering. This is a challenging situation that 
persists with open questions. Nevertheless, the theoretical approach developed 
in Refs.~\onlinecite{IzrMkrRnd-PRB_2006,RndIzrMkr-PRB-2007,
RndIzrMkr-MMET-2008} allows one to determine intervals of parameters in which 
only one of the above mechanisms plays the main role, i.e., it is possible to know 
when the channels can be considered as two independent one-dimensional 
conductors and when they exhibit a strong interaction. Moreover, that theoretical 
approach indicates a rather small region of transition between the above 
possibilities, where analytical results does not exist. The whole problem thus split 
off can be addressed by two complementary approaches. When the intra-mode 
scattering prevails the transport through the waveguide can be described by the 
theory proposed for the SRB waveguide but with the proper scaling parameters. 
When the inter-mode scattering prevails, it is necessary to apply the special 
approach developed here. Our results shows different transport regimes from 
localization to the ballistic, with the possibility, depending upon the waveguide 
parameters, of finding the coexistence and diffusive regimes. The coexistence 
regime can be present because of the hierarchy of mode-attenuation lengths 
associated with the square-gradient-scattering mechanism. Also, it is shown that 
the role of gradient mechanism of scattering is decisive, giving rise to a quite 
unexpected result. Specifically, it is demonstrated that in this case the partial 
transmissions along two channels are practically equal, and this effect is entirely 
due to the influence of the gradient terms in the expression for partial attenuation 
lengths. Due to the gradient-scattering mechanism that leads to the strong 
interaction between channels, the diffusive regime can take place.

Our results can help to understand specific properties of surface scattering in 
waveguides and electronic devices, in dependence on correlation properties of 
scattering profiles. In particular, one can understand the conditions under which 
the square-gradient mechanism of scattering has to be taken into account.

%%%%%%%%%%%%%%%%%%%%%%%%%%%%%%%%%%%%%%%%%%%%%%%%%%%%%%%%%%%%%%%%%%%%%%%%%%%%%%%
%merlin.mbs 2010-03-15 4.21a (PWD, AO, DPC)

\end{document}